\documentclass{article}

\usepackage{authblk}
\usepackage{arxiv}
\usepackage{caption}
\usepackage[utf8]{inputenc} 
\usepackage[T1]{fontenc}    
\usepackage{hyperref}       
\usepackage{url}            
\usepackage{booktabs}       
\usepackage{amsfonts}       
\usepackage{nicefrac}       
\usepackage{microtype}      
\usepackage{lipsum}
\usepackage{enumerate}
\usepackage{subfigure}
\usepackage{epstopdf}
\usepackage{tikz}
\usepackage{url}
\usepackage{pdfpages}
\usepackage{afterpage}
\usepackage{color, colortbl}
\usepackage{amsmath}
\usepackage{footmisc}
\usepackage{stfloats}
\usepackage{etoolbox}
\usepackage{listings}
\usepackage[utf8]{inputenc}
\usepackage{graphicx}
\usepackage{booktabs}
\usepackage{multirow}
\usepackage[flushleft]{threeparttable}
\usepackage{float}
\usepackage{verbatim}
\usepackage{hyperref}
\def\BibTeX{{\rm B\kern-.05em{\sc i\kern-.025em b}\kern-.08em
    T\kern-.1667em\lower.7ex\hbox{E}\kern-.125emX}}
\title{Benchmarking the Performance and Energy Efficiency of AI Accelerators for AI Training}
\author{Yuxin Wang}
\author{Qiang Wang}
\author{Shaohuai Shi}
\author{Xin He}
\author{Zhenheng Tang}
\author{Kaiyong Zhao}
\author{Xiaowen Chu \thanks{\{yxwang, qiangwang, csshshi, csxinhe, zhtang, kyzhao, chxw\}@comp.hkbu.edu.hk}}
\affil{High Performance Machine Learning Lab \\ Department of Computer Science \\ Hong Kong Baptist University}

\begin{document}

\maketitle

\begin{abstract}
Deep learning has become widely used in complex AI applications. Yet, training a deep neural network (DNNs) model requires a considerable amount of calculations, long running time, and much energy. Nowadays, many-core AI accelerators (e.g., GPUs and TPUs) are designed to improve the performance of AI training. However, processors from different vendors perform dissimilarly in terms of performance and energy consumption. To investigate the differences among several popular off-the-shelf processors (i.e., Intel CPU, NVIDIA GPU, AMD GPU, and Google TPU) in training DNNs, we carry out a comprehensive empirical study on the performance and energy efficiency of these processors \footnote{The benchmark results include performance of CPU, GPUs, TPUs; and energy of GPUs.} by benchmarking a representative set of deep learning workloads, including computation-intensive operations, classical convolutional neural networks (CNNs), recurrent neural networks (LSTM), Deep Speech 2, and Transformer. Different from the existing end-to-end benchmarks which only present the training time, We try to investigate the impact of hardware, vendor's software library, and deep learning framework on the performance and energy consumption of AI training. Our evaluation methods and results not only provide an informative guide for end users to select proper AI accelerators, but also expose some opportunities for the hardware vendors to improve their software library.
\end{abstract}

\keywords{AI Accelerator, Deep Learning, CPU, GPU, TPU, Computation-intensive Operations, Convolution Neural Networks, Recurrent Neural Networks, Transformer, Deep Speech 2}

\section{Introduction}
Recent years have witnessed the fast development of deep neural networks (DNNs) \cite{lecun2015deep} in many AI applications, such as image recognition \cite{he2015deep}, object detection \cite{redmon2016you}, speech to text tasks \cite{amodei2016deep}, etc. However, training DNN models requires a considerable amount of computational resources that usually involve Graphics Processing Units (GPUs) \cite{lecun2015deep,lsddn}.

Graphics Processing Units (GPUs) \cite{luebke2006gpgpu} serve as one of the most popular accelerators for efficient DNN training. Different from conventional CPUs, a GPU is typically equipped with thousands of cores and large memory bandwidth \cite{luebke2006gpgpu}, which significantly accelerates the training and reasoning speed of DNNs.
Since 2016, the new generation of computing device - the TPU (Tensor Processing Units) \cite{jouppi2017datacenter}, has been launched by Google with better performance than many other AI accelerators, followed by Cloud TPU v2 and Cloud TPU v3 \cite{ying2018image} with improved performance and memory capacity. With its extraordinary parallel computing capability, the cloud service of TPU greatly facilitated research and applications of artificial intelligence.

Meanwhile, the development of optimized software keeps pace with advancement of the hardware. For the CPU processors, there exist highly optimized libraries like MKL and MKLDNN \cite{cyphers2018intel}. For the NVIDIA GPUs, researchers and industry developed cuDNN \cite{chetlur2014cudnn}, cuBLAS, and other CUDA based libraries, enabling peak performance of GPU cards for AI training. For the AMD GPUs, ROCm \cite{rocm-power} is being actively developed for high performance deep learning. Also, for TPUs, TensorFlow \cite{abadi2015tensorflow} is highly optimized under a large development community.

However, generations of AI accelerators from different vendors have diverse performance and energy consumption. For example, the performance of throughput could be different on GPUs from NVIDIA and AMD with similar raw capacity. In terms of performance, there exist some benchmarks including software comparison \cite{harmonia2015}, hardware comparison \cite{wei2019benchmarking,coleman2017dawnbench,gao2019aibench,fathom,tbd} and a combination of software and hardware comparison \cite{shi2016benchmarking}\cite{mlperf2019} in training DNNs. In addition, vendors provide their own benchmarks to demonstrate the performance with their own highly optimized libraries or configurations; however, these results could be unfairly compared. 

For server deployment, it is vital to reduce energy consumption for DNN training in order to reduce the long-term electric bills. There exist several techniques to balance and improve the performance and energy efficiency, including characterization of performance and energy for traditional algorithms \cite{eppminer}; dynamic voltage and frequency scaling (DVFS) techniques \cite{mei2017survey,wang2018gpgpu,tang2019impact}; job scheduling algorithms \cite{chau2017energy} for saving energy while preserving the computing efficiency of tasks.

In summary, existing benchmarks consider either only performance or energy for particular accelerators and algorithms. Furthermore, there is little study on AMD GPUs, while AMD researchers have also developed a deep learning ecosystem ROCm for users. An overview of recent AI benchmark comparison can be found in Table \ref{tab:benchmarkcompare}. 
In this paper, we make wide benchmarks on many popular AI accelerators, including the Intel CPU, NVIDIA GPUs, the AMD GPU, and Google TPUs in terms of performance and energy efficiency. Our work covers multiple workloads ranging from computation-intensive operations to real-world tasks, including image recognition, speech recognition, and translation tasks.
To make the evaluation thorough, we first evaluate the performance on low-level operations on the above aforementioned accelerators, and then we evaluate the performance and energy consumption of the end-to-end DNN training from different AI areas, including CNNs \cite{simonyan2014very}, \cite{he2015deep},  \cite{szegedy2016rethinking}, LSTM \cite{gers1999learning}, Deep Speech 2 \cite{amodei2016deep} and Transformer \cite{vaswani2017attention}. 

Our major findings are shown in Table \ref{tab:mainfindings}. On the one hand, our evaluation results provide a guide for end-users on how to choose proper AI accelerators to train their own DNN models under different considerations. For example, end-users can compare the budgets of cloud-based GPUs and TPUs for a specific model, and choose a cheaper one to train the model. On the other hand, the problems revealed in the evaluation results could be helpful for hardware or software design for further optimization. For example, GPU library engineers can have an insight into the performance with the roofline model \cite{Williams2008RooflineAI} and kernel-level optimization of better GPU utilization. The results of energy efficiency can redound to develop job scheduling algorithms for energy conservation \cite{chau2017energy}\cite{mei2017energy}, in which one should expect the task to be finished in expected time (related to performance) with less power (related to energy).

\begin{table}[htbp]
  \centering
  \caption{AI Benchmark Comparison}
    \setlength{\tabcolsep}{0.3mm}
  \scalebox{0.9}{
    \begin{tabular}{|r|lllllll|}
    \toprule
    \multicolumn{1}{|c|}{Training} & \multicolumn{1}{c|}{Benchmarks} & \multicolumn{1}{c|}{Our work} & \multicolumn{1}{c|}{MLPerf} & \multicolumn{1}{c|}{DAWNBench} & \multicolumn{1}{c|}{AIBench} & \multicolumn{1}{c|}{Fathom} & \multicolumn{1}{c|}{TBD} \\
    \midrule
          & \multicolumn{1}{c|}{CPU} & \multicolumn{1}{c|}{$\checkmark$} & \multicolumn{1}{c|}{$\checkmark$} & \multicolumn{1}{c|}{$\checkmark$} & \multicolumn{1}{c|}{$\checkmark$} & \multicolumn{1}{c|}{$\checkmark$} & \multicolumn{1}{c|}{$\checkmark$} \\
\cmidrule{2-2}    \multicolumn{1}{|c|}{AI} & \multicolumn{1}{c|}{NVIDIA GPU} & \multicolumn{1}{c|}{$\checkmark$} & \multicolumn{1}{c|}{$\checkmark$} & \multicolumn{1}{c|}{$\checkmark$} & \multicolumn{1}{c|}{$\checkmark$} & \multicolumn{1}{c|}{$\checkmark$} & \multicolumn{1}{c|}{$\checkmark$} \\
\cmidrule{2-2}    \multicolumn{1}{|c|}{Accelerators} & \multicolumn{1}{c|}{AMD GPU} & \multicolumn{1}{c|}{$\checkmark$} & \multicolumn{1}{c|}{x} & \multicolumn{1}{c|}{x} & \multicolumn{1}{c|}{x} & \multicolumn{1}{c|}{x} & \multicolumn{1}{c|}{x} \\
\cmidrule{2-2}          & \multicolumn{1}{c|}{TPU} & \multicolumn{1}{c|}{$\checkmark$} & \multicolumn{1}{c|}{$\checkmark$} & \multicolumn{1}{c|}{$\checkmark$} & \multicolumn{1}{c|}{x} & \multicolumn{1}{c|}{x} & \multicolumn{1}{c|}{x} \\
    \midrule
    \multicolumn{1}{|c|}{\multirow{1}[6]{*}{Metrics}} & \multicolumn{1}{c|}{PERF} & \multicolumn{1}{c|}{$\checkmark$} & \multicolumn{1}{c|}{$\checkmark$} & \multicolumn{1}{c|}{$\checkmark$} & \multicolumn{1}{c|}{$\checkmark$} & \multicolumn{1}{c|}{$\checkmark$} & \multicolumn{1}{c|}{$\checkmark$} \\
\cmidrule{2-2}          & \multicolumn{1}{c|}{NRG} & \multicolumn{1}{c|}{$\checkmark$} & \multicolumn{1}{c|}{x} & \multicolumn{1}{c|}{x} & \multicolumn{1}{c|}{x} & \multicolumn{1}{c|}{x} & \multicolumn{1}{c|}{x} \\
    \midrule
    \multicolumn{1}{|c|}{\multirow{3}[10]{*}{Workloads}} & \multicolumn{1}{c|}{IC} & \multicolumn{1}{c|}{$\checkmark$} & \multicolumn{1}{c|}{$\checkmark$} & \multicolumn{1}{c|}{$\checkmark$} & \multicolumn{1}{c|}{$\checkmark$} & \multicolumn{1}{c|}{$\checkmark$} & \multicolumn{1}{c|}{$\checkmark$} \\
\cmidrule{2-2}          & \multicolumn{1}{c|}{SR} & \multicolumn{1}{c|}{$\checkmark$} & \multicolumn{1}{c|}{$\checkmark$} & \multicolumn{1}{c|}{x} & \multicolumn{1}{c|}{$\checkmark$} & \multicolumn{1}{c|}{$\checkmark$} & \multicolumn{1}{c|}{$\checkmark$} \\
\cmidrule{2-2}          & \multicolumn{1}{c|}{TR} & \multicolumn{1}{c|}{$\checkmark$} & \multicolumn{1}{c|}{$\checkmark$} & \multicolumn{1}{c|}{x} & \multicolumn{1}{c|}{$\checkmark$} & \multicolumn{1}{c|}{$\checkmark$} & \multicolumn{1}{c|}{$\checkmark$} \\
\cmidrule{2-2}          & \multicolumn{1}{c|}{TNR} & \multicolumn{1}{c|}{$\checkmark$} & \multicolumn{1}{c|}{$\checkmark$} & \multicolumn{1}{c|}{x} & \multicolumn{1}{c|}{$\checkmark$} & \multicolumn{1}{c|}{x} & \multicolumn{1}{c|}{$\checkmark$} \\
\cmidrule{2-2}          & \multicolumn{1}{c|}{CIO} & \multicolumn{1}{c|}{$\checkmark$} & \multicolumn{1}{c|}{x} & \multicolumn{1}{c|}{x} & \multicolumn{1}{c|}{$\checkmark$} & \multicolumn{1}{c|}{$\checkmark$} & \multicolumn{1}{c|}{x} \\
    \midrule
    \multicolumn{1}{|c|}{Notes:} 
          & \multicolumn{7}{l|}{PERF: Performance. NRG: Energy.} \\
          & \multicolumn{7}{l|}{IC: Image Classification; SR: Speech Recognition.} \\
          & \multicolumn{7}{l|}{TR: Translation-Recurrent; TNR: Translation-Non-recurrent.} \\
          & \multicolumn{7}{l|}{CIO: Computation-intensive Operations.} \\
    \bottomrule
    \end{tabular}}
  \label{tab:benchmarkcompare}%
\end{table}%

\begin{table*}[htbp]
  \centering
  \caption{Summary of Main Findings about PERF (performance) and NRG (energy)}
  \scalebox{0.65}{
  \setlength{\tabcolsep}{0mm}
    \begin{tabular}{|c|c|c|l|}
    \toprule
    Section & AI Accelerator & Metric & \multicolumn{1}{c|}{Main Findings} \\
    \midrule
    \midrule
    \ref{subsec:opscpu} &       & \multirow{17}[12]{*}{PERF} & 1.1 Powered by AVX optimization, the CPU utilization scales linearly by threads and FLOPs of computation-intensive operations. \\
          & CPU   &       & 1.2 Matrix multiplication is better optimized than 2d convolution with regards to the FLOPS utilization. \\
\cmidrule{1-1}\cmidrule{4-4}    \ref{subsubsec:perfend2endcpu} &       &       & 1.3 The parallel computation intensity and execution pattern, peak FLOPS, and data-fetching are three facts that bottleneck the training performance. \\
\cmidrule{1-2}\cmidrule{4-4}          &       &       & 2.1 GPUs from NVIDIA and AMD perform the same the maximum utilization of peak FLOPS with high-throughput kernels. \\
    \ref{subsec:opsgpu} &       &       & 2.2 There exists limitations on the utilization of Tensor Core. the GPU utilization with Tensor Core is 9\% - 31\% belows that of the FP32 core \\
          &       &       & under the tested computation intensity. \\
\cmidrule{1-1}\cmidrule{4-4}          &       &       & 2.3 The model property (including size, FLOPs, kernel type, parameters, etc.) casts comprehensive influence on the effects of mini-batch size \\
          & GPU   &       & during DNN training, determining the highest throughput. \\
    \ref{subsec:end2endgpuperf} &       &       & 2.4 The actual utilization of Tensor Core is low during end-to-end Training, ranging from 3\% -9\%. Compared with FP32, 2$\times$ is the maximum \\
          &       &       & performance improvement by Tensor Core on the CNN. Meanwhile, the benefits of Tensor Core on NLP models are small. \\
          &       &       & 2.5 The Titan X(Pascal) prevails Radeon VII in training performance with inferior hardware configuration and superior software optimization.  \\
          &       &       & The software supporting of NVIDIA GPUs also prevails. \\
\cmidrule{1-2}\cmidrule{4-4}    \ref{subsec:opsgpu} &       &       & 3.1 The TPU v2 can reach peak FLOPS with the TensorFlow implementation of the maximum computation intensity in the test. Operations  \\
          &       &       & with larger sizes benefits more from pulsation matrix calculation inside of the TPU. \\
\cmidrule{1-1}\cmidrule{4-4}          & TPU   &       & 3.2 From TPU v2 to TPU v3, the peak FLOPS updates 2.3$\times$, whereas the memory bandwidth updates only 1.5$\times$. The update ratios lead to an  \\
    \ref{subsubsec:end2endtpuvsgpu} &       &       & unbalance in the performance improvement among DNNs with different parameters and FLOPs. \\
          &       &       & 3.3 The benefit of performance by swtching from V100 to TPU is 1.1$\times$-1.7$\times$ on the non-recurrent translation model and 1.5$\times$-2.7$\times$ on CNNs.  \\
    \midrule
    \midrule
          &       & \multirow{5}[2]{*}{NRG} & 4.1 Server level GPUs, including P100 and V100 demostrate lower energy consumption. Especially, the usage of Tensor Core make the \\
          &       &       & energy consumption lower for 2$\times$. V100 has at most 5.2$\times$ lower energy consumption than other GPUs. \\
    \ref{subsec:powerend2end} & GPU   &       & 4.2 The energy consumption benefits from larger batch size in most cases, if only the performance has about the same benefits. \\
          &       &       & 4.3 The power management of the AMD GPU is inferior to that of the NVIDIA GPU, especially for recurrent models. \\
          &       &       & 4.4 Among NVIDIA GPUs, Titan X(Pascal) has the largest energy consumption for all end-to-end DNNs. \\
    \bottomrule
    \end{tabular}}
  \label{tab:mainfindings}%
\end{table*}%

The rest of this paper is organized as follows. Section \ref{sec:BM} introduces some background knowledge related to DNNs, AI accelerators, as well as training pipeline and algorithms. Section \ref{sec:method} describes our experimental designs and setups, including hardware configurations and DNNs. Section \ref{sec:results} demonstrates our experimental results and analysis of AI accelerators with different training tasks. Related benchmarks are introduced in Section \ref{sec:relatedwork}. We finally conclude the paper in Section \ref{sec:cc}.

\section{\textbf{Preliminaries}}\label{sec:BM}
\subsection{\textbf{Deep Models}}
\begin{figure*}[htbp]
 	\subfigure[VGG16 \cite{simonyan2014very}.]
 	{\includegraphics[width=0.23\linewidth]{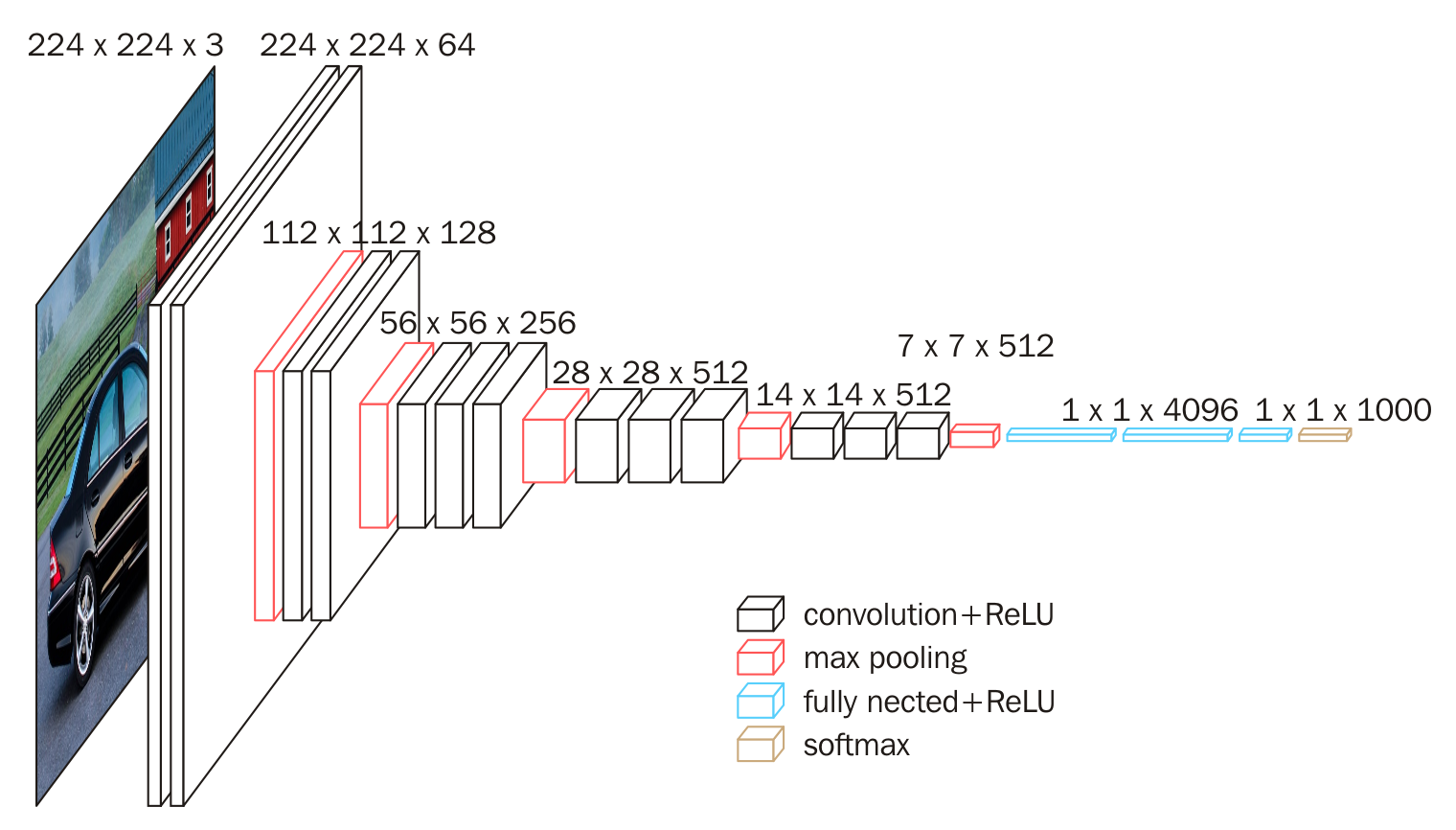}}\
 	\subfigure[LSTM \cite{gers1999learning}.]
	{\includegraphics[width=0.23\linewidth]{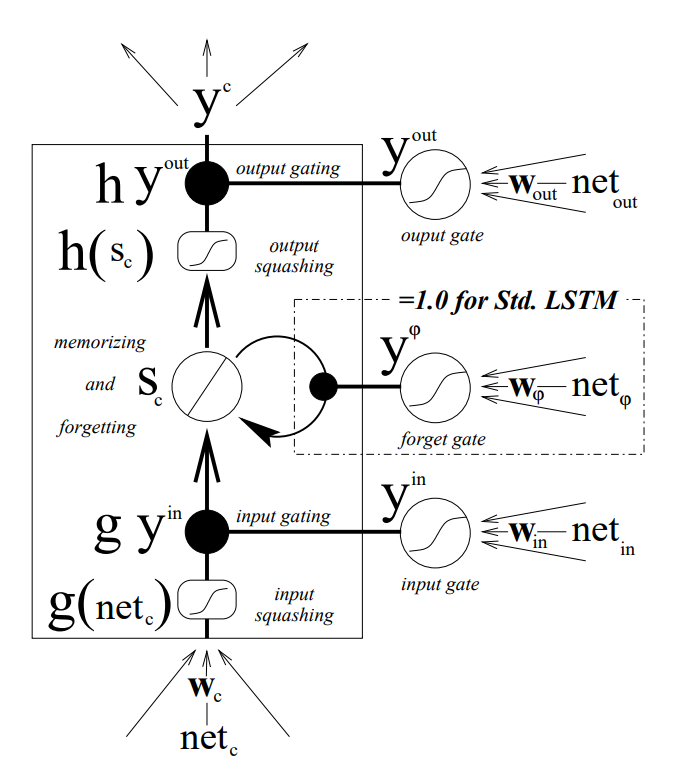}}
 	\subfigure[Deep Speech 2 \cite{amodei2016deep}.]
 	{\includegraphics[width=0.23\linewidth]{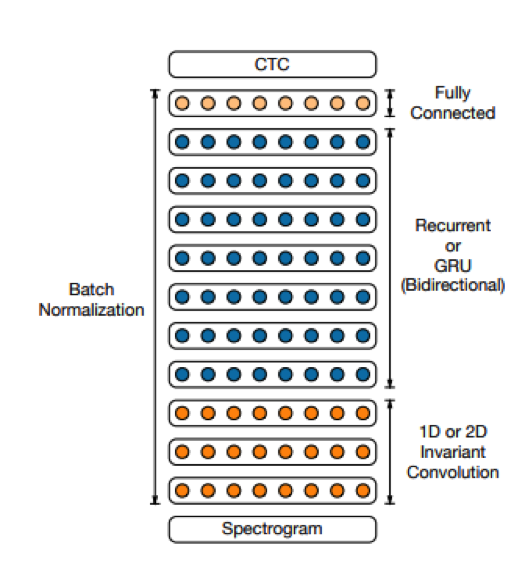}}
 	\subfigure[Transformer \cite{vaswani2017attention}.]
 	{\includegraphics[width=0.23\linewidth]{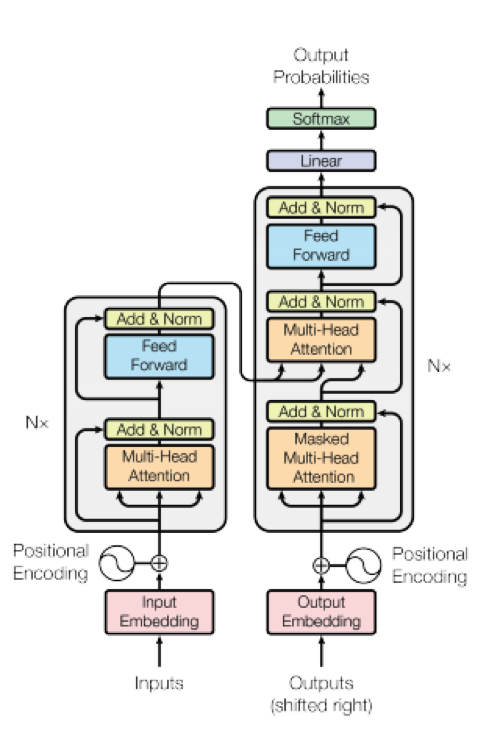}}
     \caption{Different Architecture of DNNs.}
     \label{fig:examdnns}
 \end{figure*}

In different areas of AI applications, there exist various types of deep architectures that achieve state-of-the-art results. In image classification and object detection tasks, convolutional neural networks (CNNs) are the main architectures to extract the image features automatically, among which VGG \cite{simonyan2014very}, ResNet \cite{he2015deep} and Inception \cite{szegedy2016rethinking} architectures are widely used. These CNNs also achieve excellent results in image classification and object detection, tasks of popular ImageNet challenge \cite{deng2009imagenet}. In the area of natural language processing(NLP), recurrent neural network (RNN) was one of the successful models with a long developing history, especially LSTM \cite{press2016using}. In recent years, Deep Speech 2 \cite{amodei2016deep} was proposed with state-of-the-art results on speech recognition tasks, and the attention-based model - Transformer \cite{vaswani2017attention} has achieved outstanding scores in machine translation tasks. Fig. \ref{fig:examdnns} shows some of the DNN architectures.

\subsection{\textbf{AI Accelerators}}

There are many newly developed AI accelerators. In this paper, we focus on the widely used processors, including CPU, GPU, and TPU. We will investigate FPGAs in the future. 

\paragraph{\textbf{CPU}}
CPUs are traditional processors that used in computers, while it was not good at doing highly parallel and computing-intensive tasks. In the era of deep learning, the main CPU vendor designs its many-core CPUs for these kinds of tasks. For example, the Intel Xeon processor \cite{regnier2004eta} is a powerful CPU with high computing FLOPS\footnote{In this paper, FLOPS is a performance metric indicated by \underline{FLO}ating \underline{P}oint Operations per \underline{S}econd, while FLOPs is a workload metric that represent the total number of \underline{FLO}ating \underline{P}oint Operation\underline{s}.} among Intel CPUs.  among Intel CPUs. The scalable processor was reported that it outperforms NVIDIA GPU in deep learning inference on the ResNet-50 model\footnote{\url{https://intel.ly/2k4Bxh2}}.

\paragraph{\textbf{NVIDIA and AMD GPUs}} GPUs are designed for highly parallel applications in terms of the number of computing cores and the memory access speed. The peak FLOPS has increased rapidly in the last ten years. In Table 2, we list the configurations of four recent GPUs, three of which from NVIDIA (Tesla V100, P100, and Titan X(Pascal)) and one from AMD (Radeon VII). It can be seen that the peak FP32 computing FLOPS is more than 10 TFLOPs, which is around 5$\times$ higher than CPUs. 
 
\begin{table}[htbp]
  \centering
  \caption{The Technical Specifications of GPUs}
  \scalebox{0.77}{
    \begin{tabular}{|c|c|c|c|c|}
    \toprule
    \multirow{2}[2]{*}{Product Name  } & \multirow{2}[2]{*}{Tesla V100} & \multirow{2}[2]{*}{Tesla P100} & \multirow{2}[2]{*}{Titan X(Pascal)} & \multirow{2}[2]{*}{Radeon VII} \\
          &       &       &       &  \\
    \midrule
    \midrule
    GPU  & GV100 & GP100 & GP108 & Vega 20 \\
    \midrule
    GPU Cores & 5120  & 3584  & 3584  & 3840 \\
    \midrule
    Tensor Cores & 640   & -     & -     & - \\
    \midrule
    Core Clock & 1245 MHz & 1190 MHz & 1417MHz & 1400 MHz \\
    \midrule
    Boost Clock & 1380MHz & 1329 MHz & 1531 MHz & 1750 MHz \\
    \midrule
    Memory Clock & 877 MHz & 715 MHz & 1251MHz &1000 MHz \\
    \midrule
    Memory Bus Width & 4096 bit & 4096 bit& 384 bit & 4096 bit \\
    \midrule
    Memory Bandwidth & 897.0 GB/s & 732.2 GB/s& 480.4 GB/s &  1 TB/s \\
    \midrule
    Memory Type & HBM2 & HBM2 & GDDR5X & HBM2 \\
    \midrule
    FP16 Computing & 28.26 TFLOPS & 19.05 TFLOPS & - & 26.88 TFLOPS \\
    \midrule
    FP32 Computing & 14.13 TFLOPS & 9.526 TFLOPS & 10.97 TFLOPS & 13.44 TFLOPS \\
    \midrule
    TDP & 250w & 250w & 250w & 295w \\
    \bottomrule
    \end{tabular}%
   }
  \label{tab:addlabel}%
\end{table}%

\paragraph{\textbf{Google TPUs}}

\begin{figure}[htbp]
     \centering
     \includegraphics[width=\textwidth]{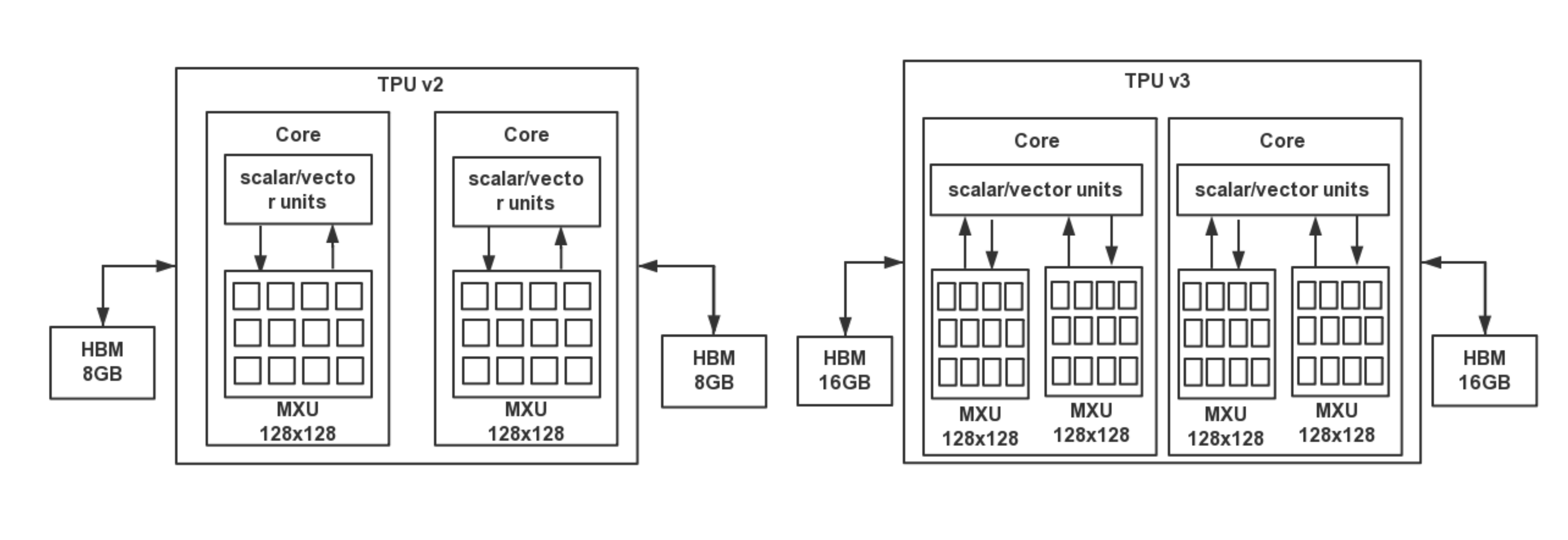}
     \caption{The structures of TPU v2 (left) and TPU v3 (right).}
     \label{fig:tpu_arch}
 \end{figure}

Tensor Processing Units (TPUs) are Google's custom-designed machine learning application-specific integrated circuits (ASICs). Each TPU device has 4 chips and each consists of 2 cores, so a TPU device contains 8 cores. Each core has scalar, vector and matrix units (MXU) and is connected with the on-chip high bandwidth memory (HBM). As shown in Fig. \ref{fig:tpu_arch}, there are two types of TPUs: TPU v2 and TPU v3. For TPU v2, the amount of HBM of each core is 8 GB. Especially, one MXU is allocated to a core. While for TPU v3, each core has two MXUs and is connected with 16 GB of HBM. TPUs support the bfloat16 format which has a wider range of values than float16 with the same 16-bit storage. TPU v2 with 8 cores (TPU v2-8) and TPU v3 with 8 cores (TPU v3-8) have peak bfloat16 computing capacity of 180 Tera bfloat16 per second and 420 Tera bfloat16 per second respectively.
 Additionally, TPU v2 Pod is assembled by 64 TPU v2 devices, containing 512 TPU v2 cores. TPU v3 Pod provides a maximum of 256 TPU v3 devices and consists of a total 2048 TPU v3 cores. 

\begin{figure*}[htbp]
     \centering
     \includegraphics[width=0.8\textwidth]{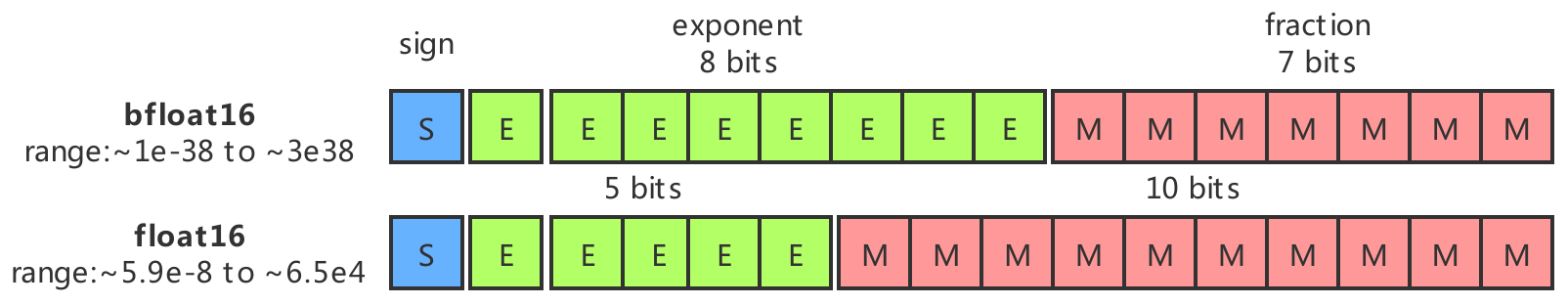}
     \caption{The comparison between bfloat16 and float16.}
    \label{fig:bfloat}
 \end{figure*}

 \subparagraph{bfloat16 and float16}The MXU in each TPU core is used to execute 16K multiply-accumulate operations in each cycle. Besides, MXU supports mixed precision training, i.e. its input and output are 32-bit floating point values and it can use bfloat16 for activation and gradients. Compared with IEEE half-precision floating point (fp16), bfloat16 has a wider range of values because it has one sign bit, eight exponent bits, and seven mantissa bits plus one implicit mantissa bit, as shown in Fig. \ref{fig:bfloat}. Using bfloat16 can help reduce the size of data in memory, making larger models available for the same size of memory, while ensuring no degradation of converged accuracy.

\subsection{\textbf{Training Pipeline}}

\paragraph{\textbf{Mini-batch SGD} }

\begin{figure}[htbp]
	\centering     
		\includegraphics[width=0.65\linewidth]{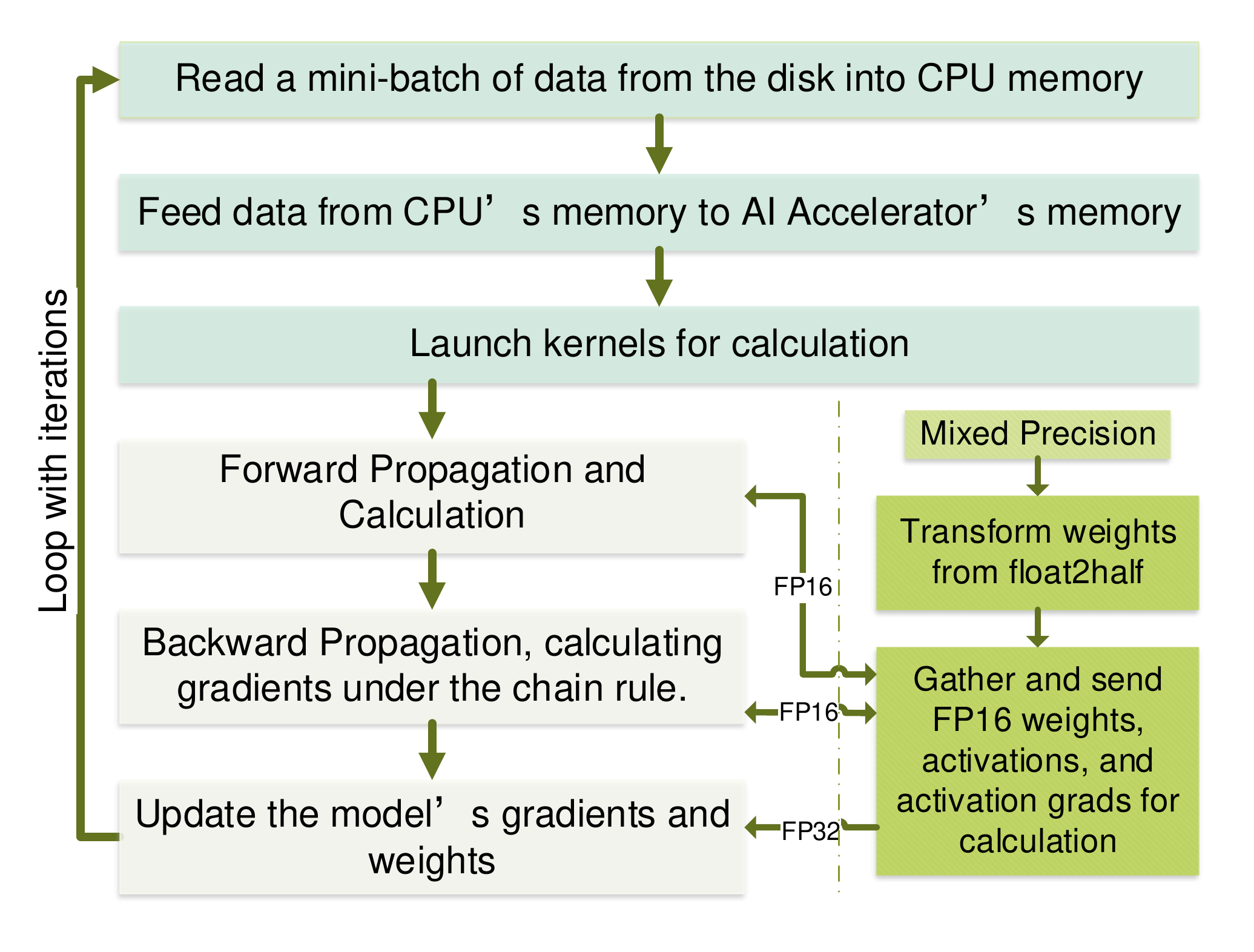}
		\caption{The training pipeline of deep neural networks. Note that FP16 will be replaced by bfloat16 when training DNNs on TPUs.}
		\vspace{-1.2 em}
		\label{fig:sgd}
\end{figure}

Mini-batch SGD \cite{li2014efficient} is a derivative algorithm of the Stochastic Gradient Descent method (SGD) , which divides the entire data set into multiple subsets, and iteratively update the model parameters according to the first-order gradients at current mini-batch of data. The training process during a single iteration can be divided into the following steps. As is shown in Fig. \ref{fig:sgd}, a single iteration starts with the process of reading data from the computer's disk to the CPU's memory, and it ends with updates of parameters. The training process is to repeat the iteration until some terminating criteria. We generally use the average iteration time to measure the performance of training on particular software and hardware. 

\paragraph{\textbf{Mixed Precision Training}}
The mixed precision \footnote{The mixed precision mainly exploits FP16(on Tensor Cores)/bfloat16(on TPU cores) as computation during the forward and backward passes.} training technique is a very successful training algorithm that uses only low-bit floating points to do the computation of forward and backward during training such that the hardware resource can be better utilized. Typically, in mixed precision, FP32 master weights and loss scaling are adopted to avoid the instability that FP16/bfloat16 precision might trigger. The training process is also shown in Fig. \ref{fig:sgd}.

\section{\textbf{Methodology}}\label{sec:method}
In this section, we introduce the methodology of our evaluation for demonstrating a comparison of performance and energy among multiple accelerators. We first present the selected hardware settings and DNNs from different AI areas. Then we illustrate our evaluation methods.

\subsection{\textbf{Hardware Setup}}
As we would like to evaluate the most commonly used accelerators for DNN training, we select many-core processors from four vendors including Intel \footnote{Note that the CPU instance used in our experiments has 24 vCPUs on a rented Alibaba Cloud ECS Computing Type c5, which is, in fact, a package of 12 physical cores, serving only half of peak theoretical FLOPS: 1.92T FLOPS.}, NVIDIA, AMD, and Google. For each vendor, we select one to three processors for evaluation. The details of the selected accelerators are listed in Table \ref{tab:hardwarespecs}, which presents the key parameters related to the performance of accelerators.

\begin{table*}[!ht]
  \centering
  \caption{Hardware Setup}
 \scalebox{0.76}{
    \begin{tabular}{|c|c|c|c|c|c|c|}
    \toprule
    Vendor & Accelerator Model & Memory & Theoretical FLOPS & Memory Bdw & Memory Type & CPU \\
    \midrule
    \midrule
    Intel & An Alibaba Cloud CPU, 12 physical cores.& 48GB  & 3.84 T(FP32)     &  119.21 GB/s & DDR4  & Xeon Platinum 8163 \\\hline
    \multirow{3}{*}{NVIDIA}& Titan X(Pascal) & 12GB  & 11 T(FP32)  & 480.4 GB/s & GDDR5X &  i7-7820X  \\
     & Tesla P100 & 16GB  & 9.5 T(FP32)  & 732.2 GB/s & HBM2  & i7-6800K   \\
     & Tesla V100 & 16GB  & 112 T(Tensor Core) & 897.0 GB/s & HBM2  & i7-6800K  \\\hline

    AMD & Radeon VII & 16GB  & 13.44 T(FP32)  & 1 TB/s & HBM2  &  i7-4790 \\\hline
    
    \multirow{2}{*}{Google} & TPU v2-8 & 64GB  & 180 T (bfloat16) &  600 GB/s & HBM   & - \\
    
     &TPU v3-8 & 128GB & 420 T (bfloat16) & 900 GB/s & HBM   & - \\
    \bottomrule
    \end{tabular}%
    }
  \label{tab:hardwarespecs}%
\end{table*}%

\subsection{\textbf{Evaluated Workloads}}
\subsubsection{\textbf{Computation-intensive Operators}}DNN models are mainly stacked by many layers, which generally invoke two main resource-consuming operators (ops), including the matrix multiplication (Matmul) and convolution in 2d dimension (Conv2d). During training, high-throughput kernels will conduct computation-intensive calculations for parallel operations like Matmul and Conv2d. Low-throughput kernels require serial operations and longer training time.

To evaluate the performance of ops, the input data are synthetic tensors of different FLOPs. For the Matmul operator, tensor dimensions range from 2048$\times$2048 to 8192$\times$8192, a typical workload of Fully Connected DNN model (For example, the Transformer model); for the Conv2d operator, inputs and filters are selected from Resnet50 under different batch sizes, including 128 and 256. To ensure the utilization of accelerators, we select to show results of small, medium, and large sizes of input tensors, which are listed in Table \ref{tab:op}. The F, K, and S in Table \ref{tab:op} refer to input size, kernel size, and strides, respectively.

\paragraph{\textbf{CUDA C++ Setting}}
We adopted the CUDA C++ codes in DeepBench \footnote{https://github.com/baidu-research/DeepBench.git} and made two major revisions to it. Firstly, we erase the unnecessary kernels launched during the training iterations to ensure the best performance of cuBLAS. Secondly, we have the forward algorithm fixed at IMPLICIT\_PRECOMP\_GEMM while input Tensors vary, which eliminates the differences in real forwarding FLOPs for ops in the experiment. 

\paragraph{\textbf{TensorFlow Setting}}
TensorFlow has provided users with the necessary tools for profiling. By printing the timeline of training processes, we get the accurate calculation time of GPU kernels. 

\begin{table}[!ht]
  \centering
  \caption{Input Tensor Sizes of Ops}
 \scalebox{0.78}{
    \begin{tabular}{|c|c|c|c|}
    \toprule
    \multirow{2}{*}{Matmul Shape} & Matmul I & Matmul II & Matmul III \\ \cmidrule{2-4}  
    & (2048, 2048) & (4096, 4096) & (8192, 8192) \\
    \midrule
    FLOPs & 1.72E+10 & 1.37E+11 & 1.10E+12 \\\hline\hline
    \multirow{3}{*}{Conv2d Shape} & Conv2d I & Conv2d II & Conv2d III \\ \cmidrule{2-4}  
    & F(256, 224, 224, 3) & F(128, 112, 112, 64) & F(256,56,56,128) \\
    \cmidrule{2-4}          
    & K(7, 7, 3, 64) S(2,2) &  K(3, 3, 64, 128) S(1,1) &  K(3, 3, 128, 256) S(1,1)\\
    \midrule
    FLOPs & 5.72E+10 & 2.28E+11 & 4.40E+11 \\
    \bottomrule
    \end{tabular}%
    }
  \label{tab:op}%
\end{table}%

\subsubsection{\textbf{DNNs}} To cover comprehensive AI applications, we choose DNN models from image classification with CNNs, language models with LSTM, speech recognition with Deep Speech 2 and the state-of-the-art Transformer language model. For CNNs, we choose ResNet-50 \cite{he2015deep}, Inception V3 \cite{szegedy2016rethinking}, and VGG16\cite{simonyan2014very} on the ImageNet \cite{deng2009imagenet} dataset. For LSTM, we selected the typical 2-Layer LSTM on the PTB dataset. For the Deep Speech 2 architecture, we train the model on the AN4 dataset. For Transformer, we test the model on the WMT14 EN-DE dataset. The details of DNN configurations are shown in Table \ref{tab:data}. 


\begin{table}[htbp]
  \centering
  \caption{Model and Dataset Setting}
    \scalebox{0.95}{
  \setlength{\tabcolsep}{0.3mm}
    \begin{tabular}{|c|lllll|}
    \toprule
    \multirow{2}[2]{*}{Networks} & \multicolumn{1}{c|}{\# Params} & \multicolumn{1}{c|}{Theoretical } & \multicolumn{1}{c|}{\multirow{2}[2]{*}{Datasets}} & \multicolumn{1}{c|}{\multirow{2}[2]{*}{\# Samples}} & \multicolumn{1}{c|}{Input } \\
          & \multicolumn{1}{c|}{(million)} & \multicolumn{1}{c|}{MACs (GFLOPs)} & \multicolumn{1}{c|}{} & \multicolumn{1}{c|}{} & \multicolumn{1}{c|}{Sizes} \\
    \midrule
    \midrule
    VGG16 & \multicolumn{1}{c|}{138} & \multicolumn{1}{c|}{86.66} & \multicolumn{1}{r|}{} & \multicolumn{1}{r|}{} & \multicolumn{1}{c|}{224*} \\
\cmidrule{1-3}    Resnet50 & \multicolumn{1}{c|}{26} & \multicolumn{1}{c|}{32.79} & \multicolumn{1}{c|}{ImageNet} & \multicolumn{1}{c|}{1.2 m} & \multicolumn{1}{c|}{224} \\
\cmidrule{1-3}    Inception V3 & \multicolumn{1}{c|}{27} & \multicolumn{1}{c|}{23.84} & \multicolumn{1}{r|}{} & \multicolumn{1}{r|}{} & \multicolumn{1}{c|}{*3} \\
    \midrule
    2-Layer LSTM & \multicolumn{1}{c|}{51} & \multicolumn{1}{c|}{6.91} & \multicolumn{1}{c|}{PTB} & \multicolumn{1}{c|}{42 k} & \multicolumn{1}{c|}{20} \\
    \midrule
    Deep Speech 2 & \multicolumn{1}{c|}{87} & \multicolumn{1}{c|}{40.99} & \multicolumn{1}{c|}{AN4} & \multicolumn{1}{c|}{948} & \multicolumn{1}{c|}{100-400} \\
    \midrule
    \multirow{2}[2]{*}{Transformer} & \multicolumn{1}{c|}{\multirow{2}[2]{*}{65}} & \multicolumn{1}{c|}{\multirow{2}[2]{*}{46.08}} & \multicolumn{1}{c|}{WMT14} & \multicolumn{1}{c|}{\multirow{2}[2]{*}{36 m}} & \multicolumn{1}{c|}{\multirow{2}[2]{*}{512}} \\
          & \multicolumn{1}{c|}{} & \multicolumn{1}{c|}{} & \multicolumn{1}{c|}{EN-DE} & \multicolumn{1}{c|}{} &  \\
    \midrule
    Note: & \multicolumn{5}{l|}{For nlp models, input sizes refer to the sequence lengths here.} \\
    \bottomrule
    \end{tabular}}
  \label{tab:data}%
\end{table}%

\subsection{\textbf{Evaluation Methods}}
\paragraph{\textbf{Evaluation Metrics}} In order to present the readers a comprehensive scope of different tasks and AI accelerators, we use performance and energy\footnote{For the CPU and TPUs, performance is the key metric for benchmarking. For GPUs, the above two metrics are adopted for analysis.} as the evaluation metrics. For the performance, 
we report the throughput of the accelerators in terms of samples per second (Samples/s). \footnote{The units in pictures of samples are images for CNNs, sentences for LSTM, utterances for Deep Speech 2, and token for Transformer, respectively.} For the energy, we sample the system power (in Watt) in every 2ms during the training process using the built-in interfaces provided by NVIDIA Management Library \cite{NVML} on NVIDIA GPUs and ROCm System Management Library \cite{rocm-power} on the AMD GPU. The energy is directly derived from the evaluated performance and power as $\frac{power}{performance}$. The metric details are defined in Table \ref{tab:metrics}.
\begin{table}[htbp]
  \centering
\caption{Definition of Metrics}
  \scalebox{0.86}{
  \setlength{\tabcolsep}{2mm}
    \begin{tabular}{|c|l|c|}
    \toprule
    Index & \multicolumn{1}{c|}{Definition} & Unit \\
    \midrule
    \midrule
    \multirow{2}[4]{*}{Performance } & Time duration for the computing device to process& \multirow{2}[4]{*}{Second} \\
\cmidrule{2-2}          &  a certain mini batch over an iteration. &  \\
    \midrule
    \multirow{2}[4]{*}{Energy} & The electrical energy cost of the computing device  & \multirow{2}[4]{*}{J per sample} \\
\cmidrule{2-2}          & to process a sample.  &  \\
    \bottomrule
    \end{tabular}%
    }
    \vspace{-1.2 em}
  \label{tab:metrics}%
\end{table}%

     

\begin{table}[htbp]
  \centering
  \caption{Software Setup}
  \scalebox{0.86}{
    \setlength{\tabcolsep}{0.55mm}
    \begin{tabular}{|c|llllll|}
    \toprule
    \multirow{2}[4]{*}{Accelerators} & \multicolumn{5}{c|}{Workloads}        & \multicolumn{1}{c|}{\multirow{2}[4]{*}{Libraries}} \\
\cmidrule{2-6}          & \multicolumn{1}{c|}{Ops} & \multicolumn{1}{c|}{Transformer} & \multicolumn{1}{c|}{CNNs} & \multicolumn{1}{c|}{LSTM} & \multicolumn{1}{c|}{Deep Speech 2} &  \\
    \midrule
    \midrule
    Intel CPU & \multicolumn{1}{c|}{\multirow{2}[4]{*}{ab}} & \multicolumn{1}{r|}{} &       &       & \multicolumn{1}{c|}{} & \multicolumn{1}{c|}{MKL-2019.4-243} \\
\cmidrule{1-1}\cmidrule{7-7}    NVIDIA GPUs & \multicolumn{1}{c|}{} & \multicolumn{1}{r|}{} &       & \multicolumn{1}{c}{c} & \multicolumn{1}{r|}{} & \multicolumn{1}{c|}{CUDA-10.0} \\
\cmidrule{1-2}\cmidrule{6-7}    AMD GPU & \multirow{2}[4]{*}{} & \multicolumn{1}{c|}{a} &       & \multicolumn{1}{r|}{} & \multicolumn{1}{r|}{} & \multicolumn{1}{c|}{ROCm-v2.4} \\
\cmidrule{1-1}\cmidrule{4-5}\cmidrule{7-7}    Google TPUs &       &       & \multicolumn{1}{r|}{} &       & \multicolumn{1}{c|}{/} & \multicolumn{1}{c|}{-} \\
    \midrule
    Software & \multicolumn{6}{l|}{a: Tensorflow 1.14.0; b: CUDA C++; c: PyTorch 1.1; /: no results yet} \\
    \bottomrule
    \end{tabular}}
  \label{tab:softwarespecs}%
\end{table}%

\paragraph{\textbf{Software Tools for Measurement}}
For the measurement of operations and DNNs, we make evaluations with the software listed in Table \ref{tab:softwarespecs}. As TPU mainly supports TensorFlow, we measure the TPU training performance with TensorFlow for the best performance. Note that we report the average data collected from the experiments. Data collected from both op level benchmarking and end-to-end training on different platforms are repeated for 1000 rounds(iterations), leaving out the 10 rounds that report the value of the highest and lowest metric.
\section{\textbf{Experimental Results}}\label{sec:results}
In this section, we present the experimental results and discussions, including the performance of low-level mathematical operators (Subsection \ref{subsec:ops}), performance of end-to-end training (Subsection \ref{subsec:perfend2end}), and energy consumption of end-to-end training (Subsection \ref{subsec:powerend2end}). 

\subsection{\textbf{Performance of Computation-intensive Operators}} \label{subsec:ops}
We evaluate the AI Accelerators on the two major operators (i.e., matrix multiplication and 2D convolution) that are computation-intensive and widely used in DNN training. We test the operators with small, medium, and large FLOPs, which represent different computation intensity for accelerators, as shown in Table \ref{tab:op}. 

\begin{figure}[htbp]
    \centering    
    \subfigure[Matmul. ]
    {\includegraphics[width=0.7\linewidth]{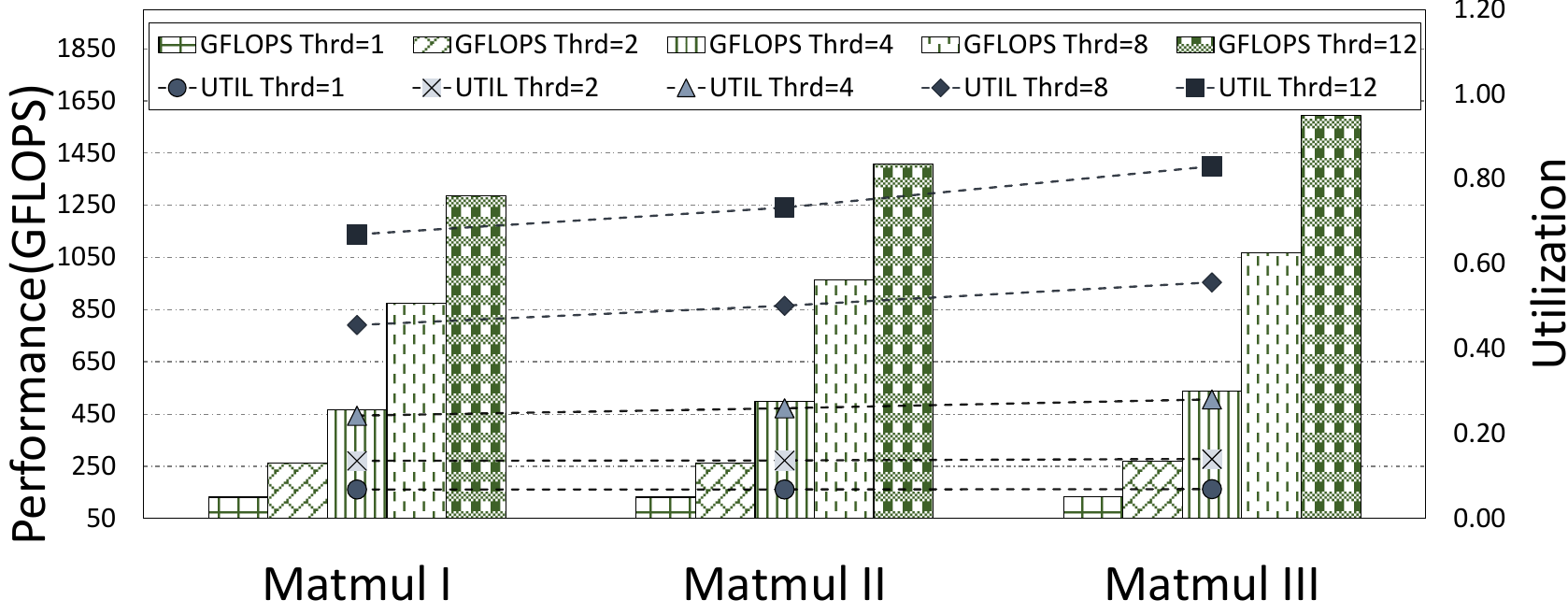}}
    \subfigure[Conv2d.]
    {\includegraphics[width=0.7\linewidth]{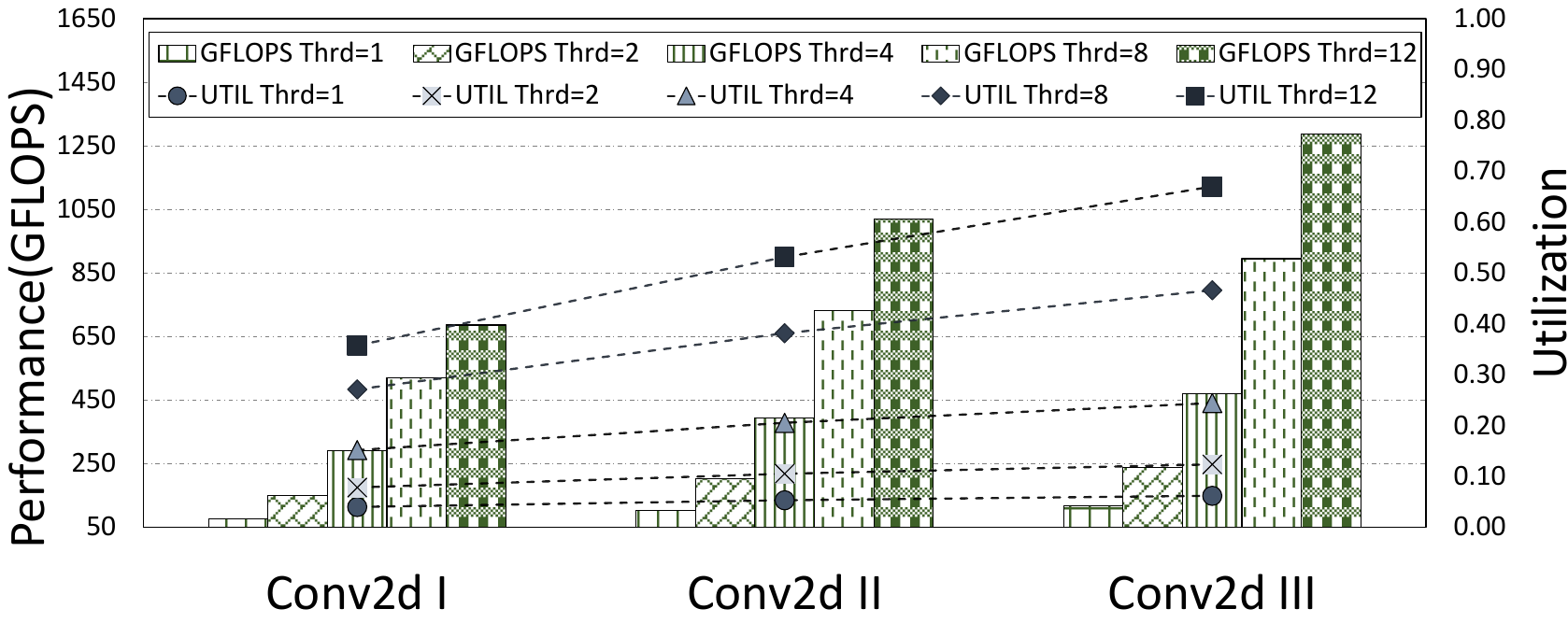}}
    \caption{The results of CPU's performance and utilization on two computation intensive operators. The lines refer to utilization, and the bars refer to TFLOPS.}
    \label{fig:opscpuresults}
\end{figure}

\subsubsection{\textbf{CPU Results}}\label{subsec:opscpu}
Multi-threading and AVX (Advanced Vector Extension) are two main techniques to exploit the many-core and SIMD capabilities of modern CPUs. To evaluate the performance of operators using an Intel Optimization version of TensorFlow with AVX512 enabled, we enumerate the number of threads from 1 to 12. The results of the utilization and performance of Intel Xeon 8163 are shown in Fig. \ref{fig:opscpuresults}. The maximum utilization is up to 83\% and 68\% on Matmul and Conv2d, respectively. For the matrix multiplication, the computing performance is almost linear to the number of threads (shown in Fig. \ref{fig:opscpuresults} as Thrd). 

Regarding the Conv2d operator, the Conv2d ops achieve overall lower utilization than Matmul ops with even higher FLOPs (Conv2d I, II compared with Matmul I, II). Since there are no memory bottlenecks for low-level operations, the results on Matmul and Conv2d indicate that there exist further opportunities to optimize the performance of Conv2d with multi-core and AVX techniques.

\subsubsection{\textbf{GPU and TPU Results}}\label{subsec:opsgpu}
\begin{figure}[htbp]
    \centering  
    \subfigure[Matmul.]
    {
        \includegraphics[width=0.7\linewidth]{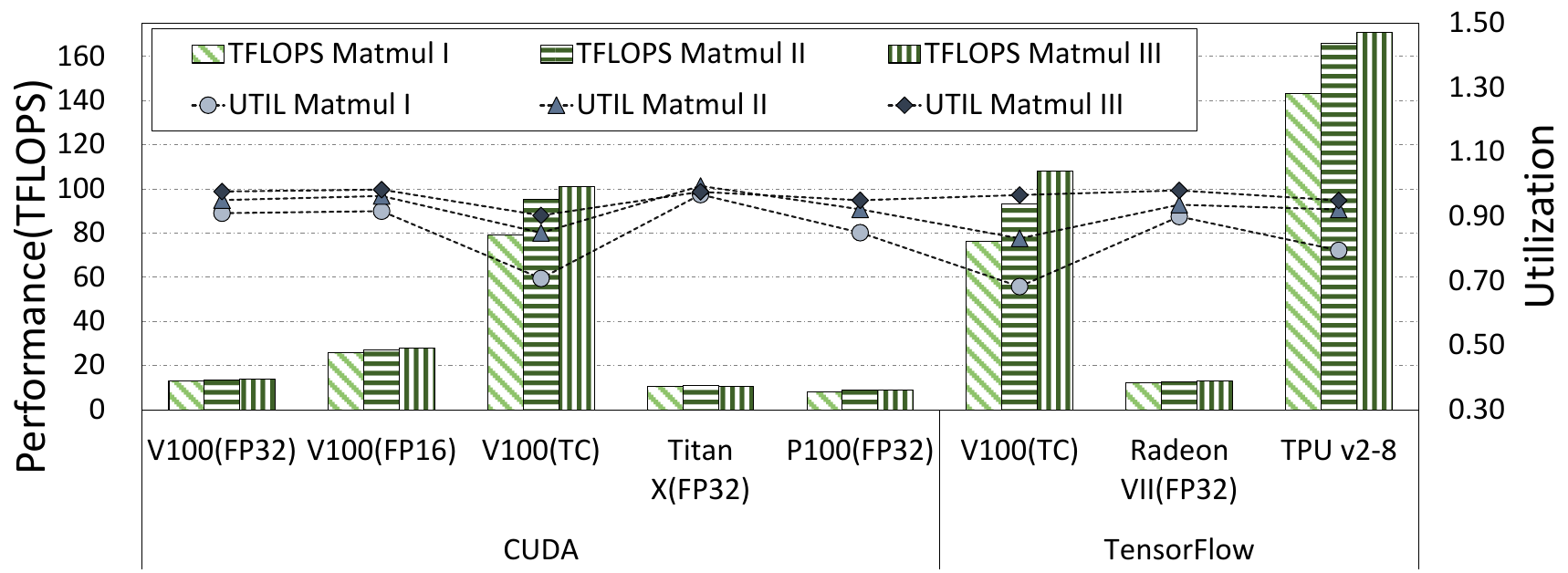}
    }
    \subfigure[Conv2d.]
    {
        \includegraphics[width=0.7\linewidth]{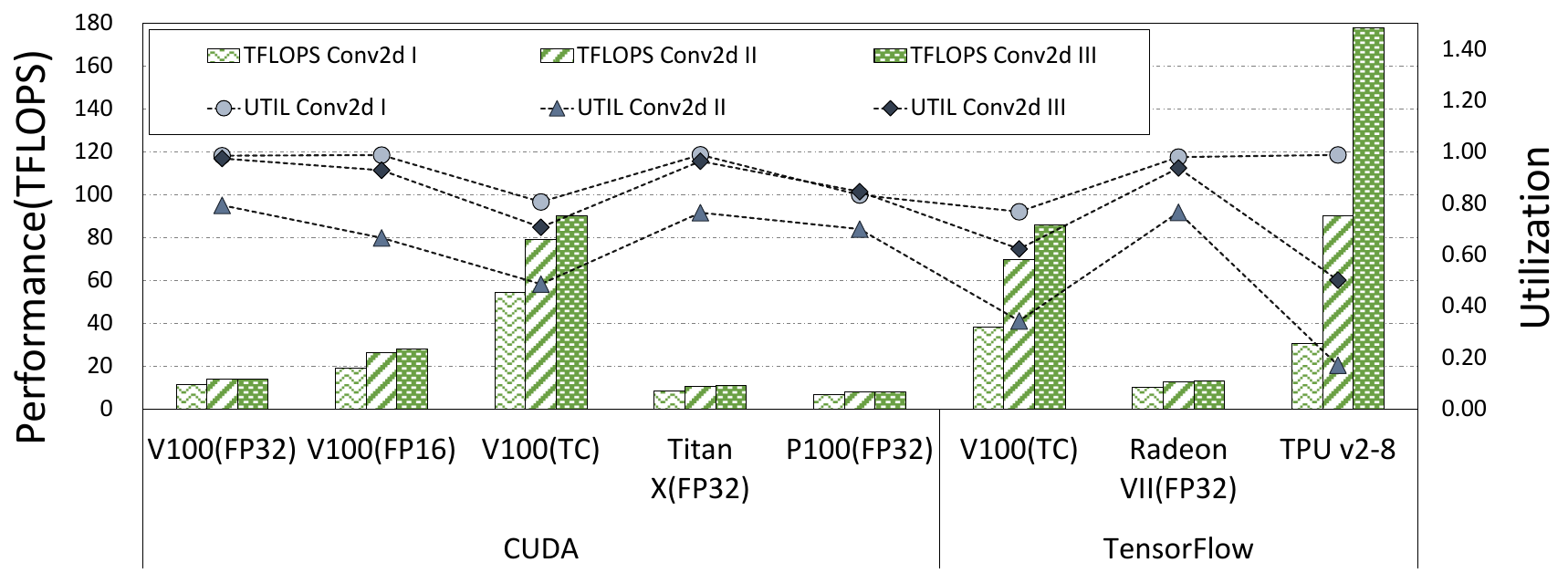}
    }
    \caption{The results of GPU and TPU's performance (FLOPS) and utilization of Matmul and Conv2d operators.}
    \label{fig:opsgpu}
\end{figure}

On the GPU and TPU, the performance and utilization of operations that have been transformed into FLOPS and percentage are shown in Fig. \ref{fig:opsgpu}. Among GPUs, P100 has the lowest utilization under all workloads. V100 has achieved up about 97$\%$ utilization for Matmul ops and 92$\%$ for Conv2d ops under FP32/FP16 Precision. The utilization of V100 with Tensor Core is 9$\%$-31$\%$ and 22$\%$-37$\%$ lower than V100 (FP32) on Matmul and Conv2d respectively (shown in Fig. \ref{fig:opsgpu} as V100 (TC) and V100 (FP32)).

Computation-intensive operations call high-throughput kernels for calculating to achieve the highest FLOPS (throughput), as can be seen from Fig. \ref{fig:opsgpu}. The gap between CUDA C++ and TensorFlow implementation of performance on Tensor Core is larger on Conv2d than Matmul. Among GPUs, V100, Titan X, and Radeon VII achieve the same best FP32 performance and utilization on both operators, which proves their ability of high-throughput calculation.
TPU V2 achieves nearly optimal utilization on both the operators, especially about 99\% on the Conv2d operator - much better than Tesla V100 with Tensor Core. We may safely draw the conclusion that there remain performance bottlenecks for V100 with Tensor Core enabled.

\subsection{\textbf{Performance of End-to-end Training}} \label{subsec:perfend2end}

\subsubsection{\textbf{CPU Results}} \label{subsubsec:perfend2endcpu}

\begin{figure}[htbp]
\centering
\includegraphics[width=0.7\linewidth]{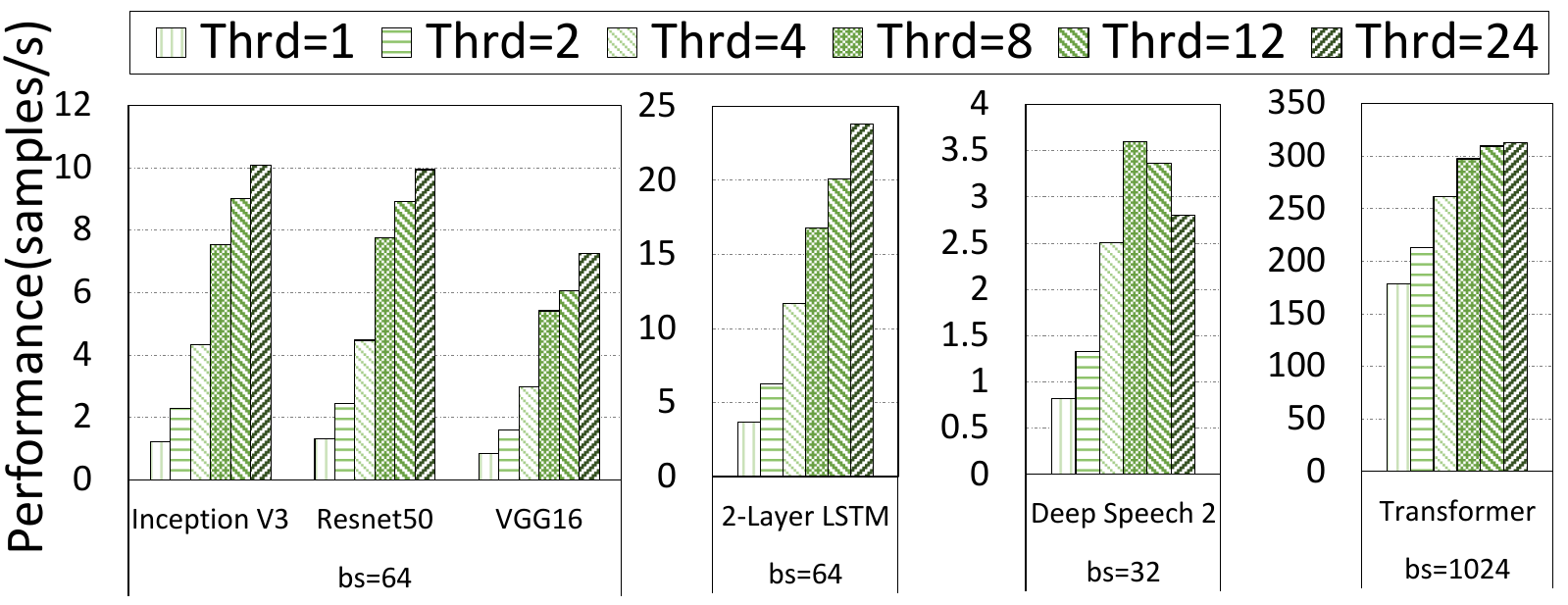}

\caption{The results of CPU's performance of end-to-end training. Note that 'bs' refers to the batch size in this paper.}

\label{fig:cpuperf-thread}
\end{figure}

The evaluated virtual CPU instance contains 12 physical cores and supports up to 8-way multiprocessing. We first evaluate how the number of threads affects training performance, which is shown in Fig. \ref{fig:cpuperf-thread}. It can be seen that on the 12-core CPU, the 2-way multiprocessing (24 threads) generally achieves the best performance on CNNs and LSTM. Transformer and Deep Speech 2 are two exceptions. When $threads=1$, the CPU utilization is high enough for Transformer, which limits the effect of multi-threading. For Deep Speech 2, the best number of threads on Deep Speech 2 of Fig. \ref{fig:cpuperf-thread} is 8. The thread for pre-processing data could lack CPU resources to do computations such that the overall performance would be even worse.

Note that with doubled number of threads (not larger than the number of physical cores), the increasing rate of performance drops due to the limitation of total CPU FLOPS and Amdahl's law - the effect of multi-threading is bottlenecked by the percentage of serial operators in the DNNs. For small threads, as we analyzed in Section \ref{subsec:ops}, the performance improvement of the parallel operator of high computation intensity is around 90\%-100\% with the double threads.

\subsubsection{\textbf{GPU Results}} \label{subsec:end2endgpuperf}

\begin{figure}[htbp]
\centering
\includegraphics[width=0.7\linewidth]{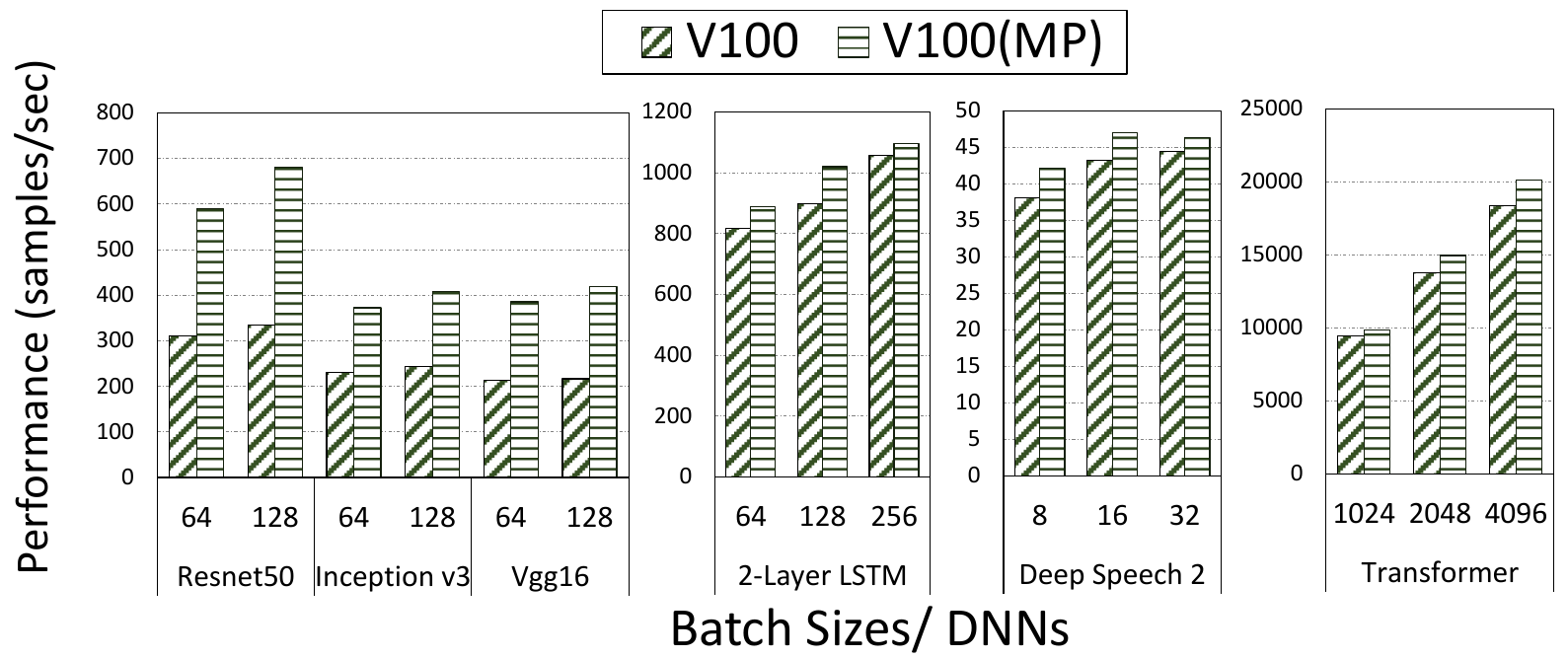}
\caption{The result of performance on V100 GPU with different mini-batch sizes using FP32 and Mixed Precision.}

\label{fig:gpuvsbatchsize}
\end{figure}

\paragraph{\textbf{Performance vs Mini-batch Size}}
There exist thousands of cores in current GPUs. When training DNNs, the GPU workload linearly increases with respect to the mini-batch size. Therefore, we first select a representative GPU (i.e., Tesla V100) in Mixed and FP32 Precision to demonstrate the relationship between performance and the mini-batch size, which is displayed in Fig. \ref{fig:gpuvsbatchsize}. As can be seen from Fig. \ref{fig:gpuvsbatchsize}, the improvement in performance by increasing batch size is greater for Mixed Precision than FP32 Precision. Smaller mini-batch sizes may not fully utilize the computing resource of GPUs. Though, the benefit of doubling the batch size from medium to the largest for Transformer and Deep Speech 2 drops in 30\% and 5\% respectively in FP32 Precision, which indicates that one should set proper mini-batch size for the particular DNN and accelerator to achieve maximum performance.

\begin{figure}[htbp]
\centering
\includegraphics[width=0.5\linewidth]{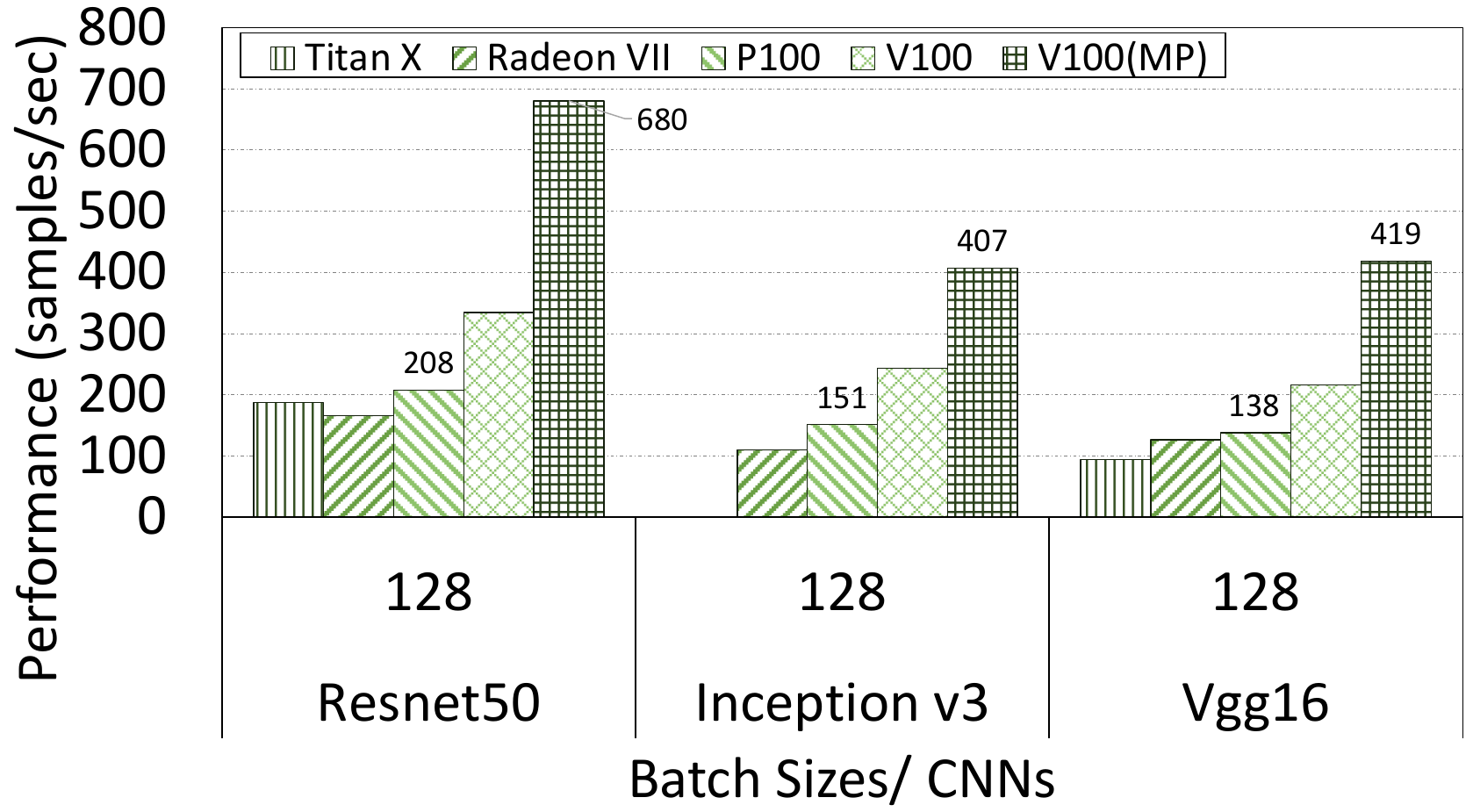}

\caption{The results of performance comparison on GPUs with CNNs. Note that the Titan X GPU runs out of memory when training inception v3 and $batch size=128$.}

\label{fig:perfvsgpus}
\end{figure}

\begin{figure}[htbp]
\centering

\includegraphics[width=0.7\linewidth]{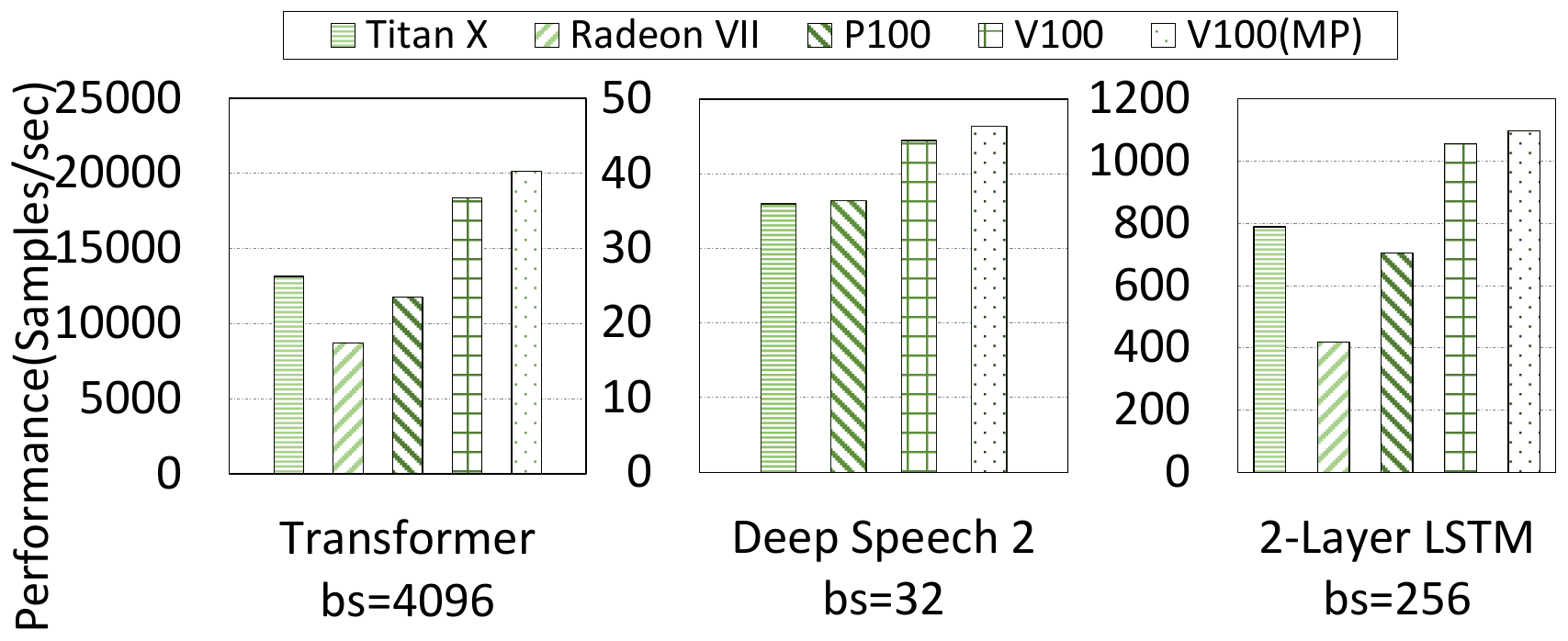}
\caption{The results of performance comparison on GPUs with NLP models. Deep Speech 2 cannot be successfully run on the Radeon VII GPU as some operators are not supported.}

\label{fig:perfnlp}
\end{figure}

The performance of end-to-end training of different DNNs on different GPUs (including NVIDIA and AMD) is shown in Fig. \ref{fig:perfvsgpus} and Fig. \ref{fig:perfnlp}. We compare their performance in two applications (i.e., Image Classification and NLP models).

\begin{table}[htbp]
  \centering
  \caption{The Average Utilization of Tensor Core}
  \scalebox{1.0}{
  \setlength{\tabcolsep}{0.5mm}
    \begin{tabular}{|c|c|c|c|c|c|c|}
    \toprule
    DNNs  & Resnet50 & Inception v3 & Vgg 16 & 2-Layer LSTM & Deep Speech 2 & Transformer \\
    \midrule
    \midrule
    Tenser Core Util & 9.2\% & 6.8\% & 7.3\% & 3.9\% & 5.1\% & 4.4\% \\
    \bottomrule
    \end{tabular}}
  \label{tab:tcutil}%
\end{table}

\paragraph{\textbf{CNNs}}

It can be seen from Fig. \ref{fig:perfvsgpus} that Tesla V100 has the best performance with both FP32 and Mixed Precision training among the tested GPUs. Under the same numerical precision, NVIDIA V100 generally achieves 1.5$\times$-2$\times$ higher performance than the AMD Radeon VII GPU. Among NVIDIA GPUs, Tesla P100 and Titan X (Pascal) have close performance as these two GPUs have similar peak FLOPS and memory bandwidth, as shown in Table \ref{tab:hardwarespecs}. When the benefit of switching from FP32 Precision to Mixed Precision is up to only 2$\times$ higher throughput for V100, the utilization of Tensor Core is 9\% at most on Resnet50 in Table \ref{tab:tcutil} - far from the peak FLOPS.
For two desktop-level GPUs between NVIDIA Titan X (Pascal) and AMD Radeon VII, the peak FLOPS, and memory bandwidth of Radeon VII is about 22\% and 52\% higher than that of Titan X (Pascal) respectively, whereas Titan X achieves slightly higher performance than Radeon VII on Resnet 50. The phenomenon indicates there exist opportunities for future optimizations on the AMD software ecosystem to better utilize the hardware ability in various deep learning workloads.

\paragraph{\textbf{NLP Models}}
The result\footnote{We present the results with the largest batch size to explore the best performance. As the connectionist temporal classification (CTC) loss function of the Deep Speech 2 is not supported on the AMD GPU, we exclude the AMD result for this case.} of performance of NLP models are shown in Fig. \ref{fig:perfnlp}. NVIDIA GPUs achieve overall higher performance than the AMD GPU. In particular, Titan X (Pascal) is nearly 1.9$\times$ and 1.5$\times$ faster than Radeon VII on recurrent and non-recurrent models, respectively. Among NVIDIA GPUs, Tesla V100 always has the best performance; however, with even slighter improvement in Mixed Precision compared to the FP32 counterpart. The margin improvement of Tesla V100 with Mixed Precision indicates that the software library should be further optimized for NLP models.

\subsubsection{\textbf{TPU Results}}\label{subsubsec:end2endtpu}
The performance of TPUs of two generations is shown in Fig. \ref{fig:perfgpuvstpu}. On three evaluated DNNs, TPU v3-8 is about 1.5$\times$-1.7$\times$ faster than TPU v2-8. However, the peak FLOPS of TPU v3-8 is around 2.3$\times$ higher than TPU v2-8, as shown in Table \ref{tab:hardwarespecs}, which indicates that FLOPS utilization on TPU v3-8 is much lower than that on TPU v2-8 and partially bottlenecked by the 1.5$\times$ improvement in memory bandwidth. The experimental results indicate that there still exist hard-ware level optimization for better performance of TPU v3-8 to fit modern deep learning networks.

\subsubsection{\textbf{Discussion}}\label{subsubsec:end2endtpuvsgpu}
\begin{figure}[htbp]
\centering

\includegraphics[width=0.57\linewidth]{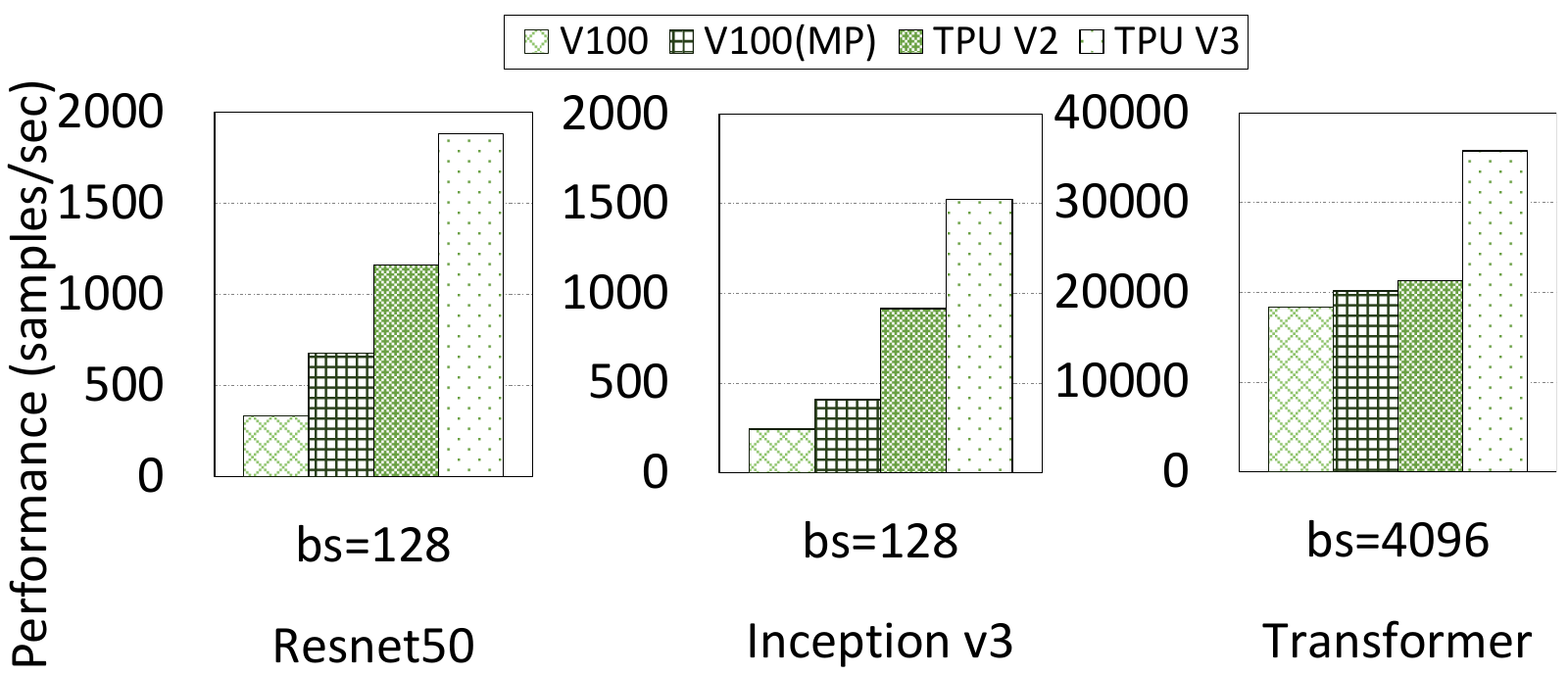}
\caption{The results of performance comparison between Tesla V100 GPU and TPUs.}
\label{fig:perfgpuvstpu}
\end{figure}

\paragraph{\textbf{Comparison Between GPU and TPU}}
V100 and TPU represent two of the fastest AI accelerators in the world. It can be seen from Fig. \ref{fig:perfgpuvstpu} that TPUs outperform Tesla V100 GPU in the three evaluated models. TPU V2 and TPU v3 have 1.6$\times$ and 3.75$\times$ more FLOPS, but 0.66$\times$ and the same memory bandwidth compared with V100 respectively. On CNNs (ResNet-50 and Inception v3), TPU V2-8 and TPU V3-8 achieve more than 1.5$\times$ and 2.7$\times$ higher throughput than Tesla V100 with Tensor Cores respectively. However, on the Transformer architecture, TPU V2-8 is very close to Tesla V100 GPU, and TPU V3-8 achieves around 1.7$\times$ faster than Tesla V100.

\paragraph{\textbf{The Limitation of Tensor Core}}
The Volta architecture transforms the convolutional parts in a neural network into matrix calculations before Tensor Cores conduct FP16 precision parallel calculations. 
Tesla V100(PCIe), in end-to-end Mixed Precision training, outperforms itself in FP32 training over two times with only 9\% utilization of the peak FLOPS of Tensor Core. However, the intensive calculations in models like Resnet50 and Transformer account for over 40\% of the total calculations. There are several possible reasons to list for this mismatch: the remaining FP32 low-throughput calculations in Mixed Precision training, the inherent low utilization of Tensor Core, and the limited memory bandwidth \cite{Williams2008RooflineAI}.

\subsection{\textbf{Energy Efficiency of End-to-end Training}} \label{subsec:powerend2end}

\begin{figure}[htbp]
\centering
\includegraphics[width=0.7\linewidth]{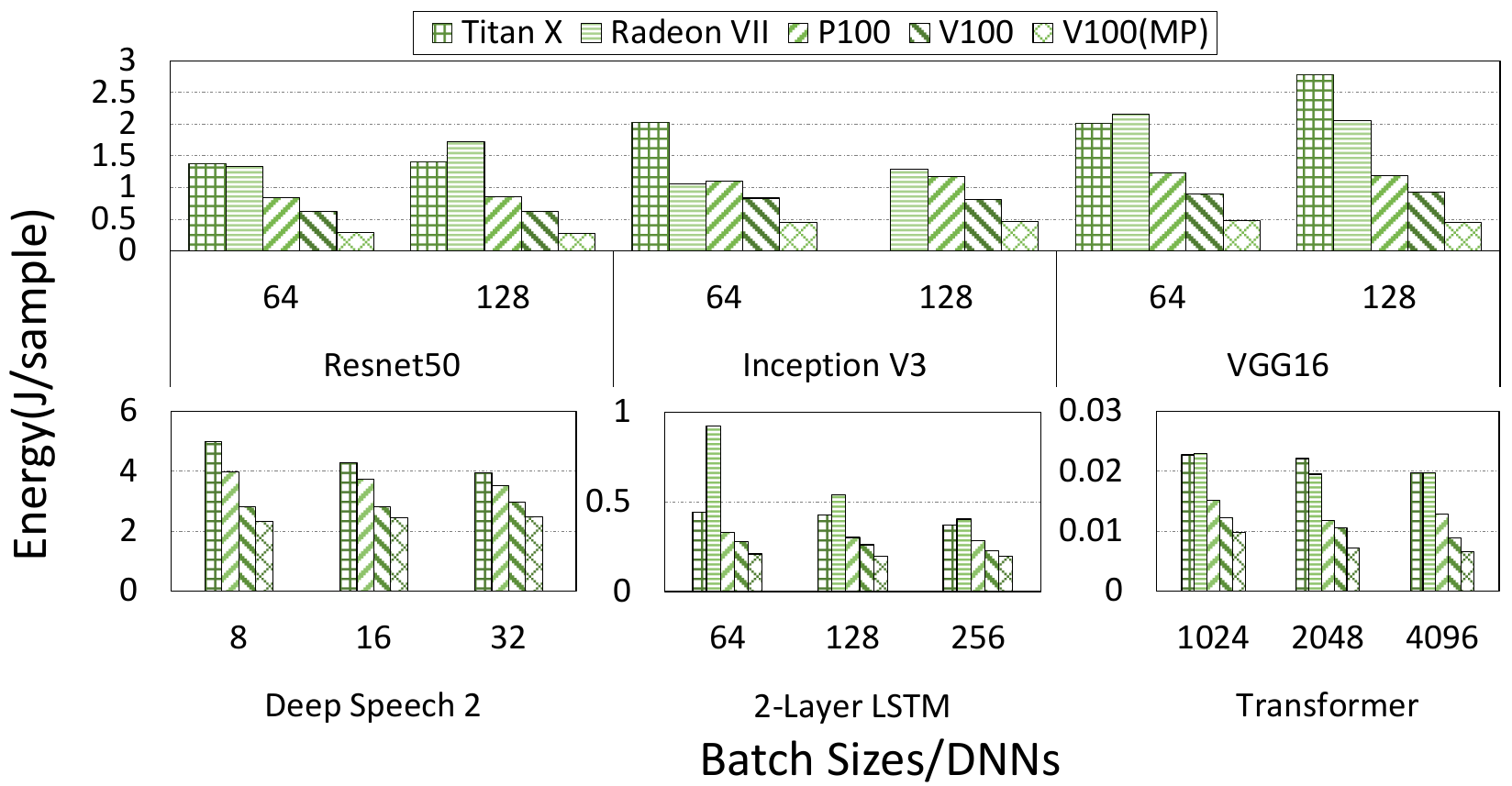}
\caption{Comparison of energy consumption on GPUs. }

\label{fig:energy}
\end{figure}

We compare the energy consumption of different GPUs among different DNN models in Fig. \ref{fig:energy}. Server-level GPUs, Tesla V100 and P100, are mostly the top-two in energy efficiency. Especially V100 has at most 5.2$\times$ lower energy consumption than other GPUs since it performs a remarkably higher training throughput with negligible power sacrifice. 
For Titan X (Pascal) and Radeon VII, increasing the processed batch size generally leads to the improvement in the resource utilization and training throughput, which consequently improves energy efficiency. 

However, for CNNs and Deep Speech 2, the larger batch size of those CNNs on P100 and V100 do not help conserve energy consumption and even lead to marginal increments, since server GPUs achieve similar training throughput on models under two batch sizes, as shown in Fig. \ref{fig:gpuvsbatchsize}. 

For recurrent models, Radeon VII should be further optimized due to high energy consumption. As can be seen from Fig. \ref{fig:energy}, the maximum energy efficiency of V100 on 2-Layer LSTM compared with Radeon VII is 4$\times$. Meanwhile, NVIDIA GPUs achieve similar energy efficiency under different batch sizes, while the Radeon VII only achieve better energy efficiency when the batch size increases.

At last, Transformer, the non-recurrent translation model, launches GPU kernels that require low training power during training. The low utilization of GPU leads to promotions in the training throughput while increasing the batch sizes, which results in better energy efficiency for all the GPUs.

For ease of reference to the experimental results, all raw numbers of end-to-end training are shown in Table \ref{tab:rawdata}.

\begin{table}[htbp]
  \centering
  \caption{End-to-end Experimental Results}
  \scalebox{0.585}{
    \begin{tabular}{|lllllllllllllllll|}
    \toprule
    \multicolumn{1}{|c|}{\multirow{2}[4]{*}{Model}} & \multicolumn{1}{c|}{DNN} & \multicolumn{2}{c|}{Resnet50} & \multicolumn{2}{c|}{Inception V3} & \multicolumn{2}{c|}{VGG16} & \multicolumn{3}{c|}{2-Layer LSTM} & \multicolumn{3}{c|}{Deep Speech 2} & \multicolumn{3}{c|}{Transformer} \\
\cmidrule{2-17}    \multicolumn{1}{|c|}{} & \multicolumn{1}{c|}{Batch Size} & \multicolumn{1}{c|}{64} & \multicolumn{1}{c|}{128} & \multicolumn{1}{c|}{64} & \multicolumn{1}{c|}{128} & \multicolumn{1}{c|}{64} & \multicolumn{1}{c|}{128} & \multicolumn{1}{c|}{64} & \multicolumn{1}{c|}{128} & \multicolumn{1}{c|}{256} & \multicolumn{1}{c|}{8} & \multicolumn{1}{c|}{16} & \multicolumn{1}{c|}{32} & \multicolumn{1}{c|}{1024} & \multicolumn{1}{c|}{2048} & \multicolumn{1}{c|}{4096} \\
    \midrule
    \multicolumn{1}{|c|}{} & \multicolumn{1}{c|}{Titan X} & \multicolumn{1}{c|}{189.25 } & \multicolumn{1}{c|}{187.43 } & \multicolumn{1}{c|}{131.25 } & \multicolumn{1}{c|}{OOM} & \multicolumn{1}{c|}{132.46 } & \multicolumn{1}{c|}{94.62 } & \multicolumn{1}{c|}{622.36 } & \multicolumn{1}{c|}{669.18 } & \multicolumn{1}{c|}{789.21 } & \multicolumn{1}{c|}{25.81 } & \multicolumn{1}{c|}{32.00 } & \multicolumn{1}{c|}{35.96 } & \multicolumn{1}{c|}{9,359.36 } & \multicolumn{1}{c|}{10,690.56 } & \multicolumn{1}{c|}{13,148.16 } \\
\cmidrule{2-17}    \multicolumn{1}{|c|}{} & \multicolumn{1}{c|}{Radeon VII} & \multicolumn{1}{c|}{206.45 } & \multicolumn{1}{c|}{166.23 } & \multicolumn{1}{c|}{133.33 } & \multicolumn{1}{c|}{110.34 } & \multicolumn{1}{c|}{125.49 } & \multicolumn{1}{c|}{126.73 } & \multicolumn{1}{c|}{145.45 } & \multicolumn{1}{c|}{261.22 } & \multicolumn{1}{c|}{419.67 } & \multicolumn{1}{c|}{-} & \multicolumn{1}{c|}{-} & \multicolumn{1}{c|}{-} & \multicolumn{1}{c|}{5,521.00 } & \multicolumn{1}{c|}{7,658.00 } & \multicolumn{1}{c|}{8,722.00 } \\
\cmidrule{2-17}    \multicolumn{1}{|c|}{} & \multicolumn{1}{c|}{P100} & \multicolumn{1}{c|}{202.47 } & \multicolumn{1}{c|}{207.61 } & \multicolumn{1}{c|}{147.69 } & \multicolumn{1}{c|}{151.23 } & \multicolumn{1}{c|}{135.58 } & \multicolumn{1}{c|}{138.12 } & \multicolumn{1}{c|}{591.25 } & \multicolumn{1}{c|}{656.29 } & \multicolumn{1}{c|}{705.28 } & \multicolumn{1}{c|}{27.59 } & \multicolumn{1}{c|}{32.00 } & \multicolumn{1}{c|}{36.36 } & \multicolumn{1}{c|}{6,973.44 } & \multicolumn{1}{c|}{11,632.64 } & \multicolumn{1}{c|}{11,796.48 } \\
\cmidrule{2-17}    \multicolumn{1}{|c|}{PERF} & \multicolumn{1}{c|}{V100} & \multicolumn{1}{c|}{311.25 } & \multicolumn{1}{c|}{334.58 } & \multicolumn{1}{c|}{230.53 } & \multicolumn{1}{c|}{243.59 } & \multicolumn{1}{c|}{213.32 } & \multicolumn{1}{c|}{216.25 } & \multicolumn{1}{c|}{817.44 } & \multicolumn{1}{c|}{898.21 } & \multicolumn{1}{c|}{1,055.87 } & \multicolumn{1}{c|}{38.10 } & \multicolumn{1}{c|}{43.24 } & \multicolumn{1}{c|}{44.44 } & \multicolumn{1}{c|}{9,431.04 } & \multicolumn{1}{c|}{13,783.04 } & \multicolumn{1}{c|}{18,350.08 } \\
\cmidrule{2-17}    \multicolumn{1}{|c|}{(samples/s)} & \multicolumn{1}{c|}{V100(MP)} & \multicolumn{1}{c|}{589.39 } & \multicolumn{1}{c|}{680.45 } & \multicolumn{1}{c|}{372.21 } & \multicolumn{1}{c|}{406.88 } & \multicolumn{1}{c|}{\textbf{385.43 }} & \multicolumn{1}{c|}{\textbf{418.59 }} & \multicolumn{1}{c|}{\textbf{889.21 }} & \multicolumn{1}{c|}{\textbf{1,019.55 }} & \multicolumn{1}{c|}{\textbf{1,096.57 }} & \multicolumn{1}{c|}{\textbf{42.11 }} & \multicolumn{1}{c|}{\textbf{47.06 }} & \multicolumn{1}{c|}{\textbf{46.38 }} & \multicolumn{1}{c|}{9,236.48 } & \multicolumn{1}{c|}{14,950.40 } & \multicolumn{1}{c|}{20,152.32 } \\
\cmidrule{2-17}    \multicolumn{1}{|c|}{} & \multicolumn{1}{c|}{TPU V2} & \multicolumn{1}{c|}{1,066.67 } & \multicolumn{1}{c|}{1,163.64 } & \multicolumn{1}{c|}{800.00 } & \multicolumn{1}{c|}{914.29 } & \multicolumn{1}{c|}{-} & \multicolumn{1}{c|}{-} & \multicolumn{1}{c|}{-} & \multicolumn{1}{c|}{-} & \multicolumn{1}{c|}{-} & \multicolumn{1}{c|}{-} & \multicolumn{1}{c|}{-} & \multicolumn{1}{c|}{-} & \multicolumn{1}{c|}{16,000.00 } & \multicolumn{1}{c|}{19,033.46 } & \multicolumn{1}{c|}{21,333.33 } \\
\cmidrule{2-17}    \multicolumn{1}{|c|}{} & \multicolumn{1}{c|}{TPU V3} & \multicolumn{1}{c|}{1,280.00 } & \multicolumn{1}{c|}{\textbf{1,882.35 }} & \multicolumn{1}{c|}{\textbf{1,488.37 }} & \multicolumn{1}{c|}{\textbf{1,523.81 }} & \multicolumn{1}{c|}{-} & \multicolumn{1}{c|}{-} & \multicolumn{1}{c|}{-} & \multicolumn{1}{c|}{-} & \multicolumn{1}{c|}{-} & \multicolumn{1}{c|}{-} & \multicolumn{1}{c|}{-} & \multicolumn{1}{c|}{-} & \multicolumn{1}{c|}{\textbf{25,600.00 }} & \multicolumn{1}{c|}{\textbf{29,383.07 }} & \multicolumn{1}{c|}{\textbf{35,772.93 }} \\
\cmidrule{2-17}    \multicolumn{1}{|c|}{} & \multicolumn{1}{c|}{CPU} & \multicolumn{1}{c|}{6.35 } & \multicolumn{1}{c|}{11.77 } & \multicolumn{1}{c|}{6.44 } & \multicolumn{1}{c|}{12.33 } & \multicolumn{1}{c|}{8.84 } & \multicolumn{1}{c|}{16.64 } & \multicolumn{1}{c|}{2.69 } & \multicolumn{1}{c|}{3.22 } & \multicolumn{1}{c|}{4.72 } & \multicolumn{1}{c|}{3.41 } & \multicolumn{1}{c|}{5.17 } & \multicolumn{1}{c|}{11.44 } & \multicolumn{1}{c|}{300.12 } & \multicolumn{1}{c|}{-} & \multicolumn{1}{c|}{-} \\
    \midrule
    \multicolumn{1}{|c|}{} & \multicolumn{1}{c|}{Titan X} & \multicolumn{1}{c|}{260.15 } & \multicolumn{1}{c|}{263.21 } & \multicolumn{1}{c|}{265.79 } & \multicolumn{1}{c|}{OOM} & \multicolumn{1}{c|}{266.72 } & \multicolumn{1}{c|}{263.15 } & \multicolumn{1}{c|}{276.08 } & \multicolumn{1}{c|}{275.25 } & \multicolumn{1}{c|}{278.21 } & \multicolumn{1}{c|}{128.74 } & \multicolumn{1}{c|}{136.58 } & \multicolumn{1}{c|}{141.55 } & \multicolumn{1}{c|}{212.64 } & \multicolumn{1}{c|}{236.35 } & \multicolumn{1}{c|}{259.23 } \\
\cmidrule{2-17}    \multicolumn{1}{|c|}{} & \multicolumn{1}{c|}{Radeon VII} & \multicolumn{1}{c|}{275.50 } & \multicolumn{1}{c|}{285.50 } & \multicolumn{1}{c|}{\textbf{141.00 }} & \multicolumn{1}{c|}{\textbf{142.00 }} & \multicolumn{1}{c|}{271.00 } & \multicolumn{1}{c|}{260.00 } & \multicolumn{1}{c|}{\textbf{134.00 }} & \multicolumn{1}{c|}{\textbf{141.00 }} & \multicolumn{1}{c|}{\textbf{169.50 }} & \multicolumn{1}{c|}{-} & \multicolumn{1}{c|}{-} & \multicolumn{1}{c|}{-} & \multicolumn{1}{c|}{126.51 } & \multicolumn{1}{c|}{149.21 } & \multicolumn{1}{c|}{172.55 } \\
\cmidrule{2-17}    \multicolumn{1}{|c|}{POW} & \multicolumn{1}{c|}{P100} & \multicolumn{1}{c|}{\textbf{169.64 }} & \multicolumn{1}{c|}{\textbf{176.51 }} & \multicolumn{1}{c|}{162.38 } & \multicolumn{1}{c|}{178.03 } & \multicolumn{1}{c|}{\textbf{166.76 }} & \multicolumn{1}{c|}{\textbf{163.45 }} & \multicolumn{1}{c|}{195.32 } & \multicolumn{1}{c|}{197.35 } & \multicolumn{1}{c|}{201.47 } & \multicolumn{1}{c|}{109.65 } & \multicolumn{1}{c|}{119.52 } & \multicolumn{1}{c|}{127.87 } & \multicolumn{1}{c|}{105.70 } & \multicolumn{1}{c|}{136.39 } & \multicolumn{1}{c|}{151.32 } \\
\cmidrule{2-17}    \multicolumn{1}{|c|}{(watt)} & \multicolumn{1}{c|}{V100} & \multicolumn{1}{c|}{193.87 } & \multicolumn{1}{c|}{206.61 } & \multicolumn{1}{c|}{191.56 } & \multicolumn{1}{c|}{198.68 } & \multicolumn{1}{c|}{190.71 } & \multicolumn{1}{c|}{201.19 } & \multicolumn{1}{c|}{226.61 } & \multicolumn{1}{c|}{234.32 } & \multicolumn{1}{c|}{240.21 } & \multicolumn{1}{c|}{107.54 } & \multicolumn{1}{c|}{121.64 } & \multicolumn{1}{c|}{131.77 } & \multicolumn{1}{c|}{115.67 } & \multicolumn{1}{c|}{144.61 } & \multicolumn{1}{c|}{162.26 } \\
\cmidrule{2-17}    \multicolumn{1}{|c|}{} & \multicolumn{1}{c|}{V100(MP)} & \multicolumn{1}{c|}{171.56 } & \multicolumn{1}{c|}{188.87 } & \multicolumn{1}{c|}{168.71 } & \multicolumn{1}{c|}{190.38 } & \multicolumn{1}{c|}{183.73 } & \multicolumn{1}{c|}{187.41 } & \multicolumn{1}{c|}{186.37 } & \multicolumn{1}{c|}{199.41 } & \multicolumn{1}{c|}{215.58 } & \multicolumn{1}{c|}{\textbf{98.21 }} & \multicolumn{1}{c|}{\textbf{115.05 }} & \multicolumn{1}{c|}{\textbf{114.77 }} & \multicolumn{1}{c|}{\textbf{90.43 }} & \multicolumn{1}{c|}{\textbf{107.04 }} & \multicolumn{1}{c|}{\textbf{132.21 }} \\
    \midrule
    \multicolumn{1}{|c|}{} & \multicolumn{1}{c|}{Titan X} & \multicolumn{1}{c|}{1.37 } & \multicolumn{1}{c|}{1.40 } & \multicolumn{1}{c|}{2.03 } & \multicolumn{1}{c|}{OOM} & \multicolumn{1}{c|}{2.01 } & \multicolumn{1}{c|}{2.78 } & \multicolumn{1}{c|}{0.44 } & \multicolumn{1}{c|}{0.41 } & \multicolumn{1}{c|}{0.35 } & \multicolumn{1}{c|}{4.99 } & \multicolumn{1}{c|}{4.27 } & \multicolumn{1}{c|}{3.94 } & \multicolumn{1}{c|}{0.0227 } & \multicolumn{1}{c|}{0.0221 } & \multicolumn{1}{c|}{0.0197 } \\
\cmidrule{2-17}    \multicolumn{1}{|c|}{} & \multicolumn{1}{c|}{Radeon VII} & \multicolumn{1}{c|}{1.33 } & \multicolumn{1}{c|}{1.72 } & \multicolumn{1}{c|}{1.06 } & \multicolumn{1}{c|}{1.29 } & \multicolumn{1}{c|}{2.16 } & \multicolumn{1}{c|}{2.05 } & \multicolumn{1}{c|}{0.92 } & \multicolumn{1}{c|}{0.54 } & \multicolumn{1}{c|}{0.40 } & \multicolumn{1}{c|}{-} & \multicolumn{1}{c|}{-} & \multicolumn{1}{c|}{-} & \multicolumn{1}{c|}{0.0229 } & \multicolumn{1}{c|}{0.0195 } & \multicolumn{1}{c|}{0.0198 } \\
\cmidrule{2-17}    \multicolumn{1}{|c|}{NRG} & \multicolumn{1}{c|}{P100} & \multicolumn{1}{c|}{0.84 } & \multicolumn{1}{c|}{0.85 } & \multicolumn{1}{c|}{1.10 } & \multicolumn{1}{c|}{1.18 } & \multicolumn{1}{c|}{1.23 } & \multicolumn{1}{c|}{1.18 } & \multicolumn{1}{c|}{0.33 } & \multicolumn{1}{c|}{0.30 } & \multicolumn{1}{c|}{0.29 } & \multicolumn{1}{c|}{3.97 } & \multicolumn{1}{c|}{3.74 } & \multicolumn{1}{c|}{3.52 } & \multicolumn{1}{c|}{0.0152 } & \multicolumn{1}{c|}{0.0117 } & \multicolumn{1}{c|}{0.0128 } \\
\cmidrule{2-17}    \multicolumn{1}{|c|}{(J/sample)} & \multicolumn{1}{c|}{V100} & \multicolumn{1}{c|}{0.62 } & \multicolumn{1}{c|}{0.62 } & \multicolumn{1}{c|}{0.83 } & \multicolumn{1}{c|}{0.82 } & \multicolumn{1}{c|}{0.89 } & \multicolumn{1}{c|}{0.93 } & \multicolumn{1}{c|}{0.28 } & \multicolumn{1}{c|}{0.26 } & \multicolumn{1}{c|}{0.23 } & \multicolumn{1}{c|}{2.82 } & \multicolumn{1}{c|}{2.81 } & \multicolumn{1}{c|}{2.96 } & \multicolumn{1}{c|}{0.0123 } & \multicolumn{1}{c|}{0.0105 } & \multicolumn{1}{c|}{0.0088 } \\
\cmidrule{2-17}    \multicolumn{1}{|c|}{} & \multicolumn{1}{c|}{V100(MP)} & \multicolumn{1}{c|}{\textbf{0.29 }} & \multicolumn{1}{c|}{\textbf{0.28 }} & \multicolumn{1}{c|}{\textbf{0.45 }} & \multicolumn{1}{c|}{\textbf{0.47 }} & \multicolumn{1}{c|}{\textbf{0.48 }} & \multicolumn{1}{c|}{\textbf{0.45 }} & \multicolumn{1}{c|}{\textbf{0.21 }} & \multicolumn{1}{c|}{\textbf{0.20 }} & \multicolumn{1}{c|}{\textbf{0.20 }} & \multicolumn{1}{c|}{\textbf{2.33 }} & \multicolumn{1}{c|}{\textbf{2.44 }} & \multicolumn{1}{c|}{\textbf{2.47 }} & \multicolumn{1}{c|}{\textbf{0.0098 }} & \multicolumn{1}{c|}{\textbf{0.0072 }} & \multicolumn{1}{c|}{\textbf{0.0066 }} \\
    \midrule
    \multicolumn{17}{|l|}{Note: '-' means the item is currently unsupported.} \\
    \bottomrule
    \end{tabular}%
    }
  \label{tab:rawdata}%
\end{table}%

\section{Related Work} \label{sec:relatedwork}
Benchmarks are key methods to make the hardware and software move forward to better targets (e.g., performance and/or energy). In the era of deep learning, training tasks are computationally intensive and resource-consuming. The running time performance and energy consumption of AI accelerators are the two major concerns for practitioners. Started from 2016, deep learning frameworks are rapidly developed to  support many types of processors and accelerators, like CPUs, GPUs, and TPUs.

Researchers \cite{harmonia2015, shi2016benchmarking} started to evaluate the performance among different deep learning frameworks and different GPUs.  However, these works mainly focused on software-level evaluation in terms of performance. Later, Stanford DAWN deep learning benchmark \cite{coleman2017dawnbench} and MLPerf \cite{mlperf2019} were developed for the evaluation of training and inference performance under different software and hardware platforms. These two open benchmark platforms have attracted many submissions from vendors, but they mainly focused on the end-to-end training performance without reasoning on the results. Several new benchmarks including ParaDnn \cite{wei2019benchmarking}, AIBench \cite{gao2019aibench}, Fathom \cite{fathom}, and TBD \cite{tbd} conduct better analysis, whereas lack comprehensive scope of AI accelerators or the energy efficiency. We list some benchmarks in Table. \ref{tab:benchmarkcompare} for comparison. Recently Wang et al. \cite{wei2019benchmarking}  proposed ParaDnn to measure the performance of various hardware including Intel CPU, NVIDIA GPU and Google TPU, which was the closest to our work. 

The energy consumption is of great importance to the servers that run resource-intensive tasks, but there is little study measuring the power and energy consumption of DNN training tasks. One related work is \cite{tang2019impact} which studied the impact of GPU dynamic voltage and frequency scaling (DVFS) on the training performance and energy. 

\section{Conclusion} \label{sec:cc}
In this paper, we made a comprehensive evaluation of the training performance and energy consumption on various modern AI accelerators including Intel CPU, AMD/NVIDIA GPUs, and Google TPUs, covering a representative set of deep learning workloads (CNNs for image recognition, recurrent and non-recurrent neural networks for translation, and Deep Speech 2 for speech recognition). Our benchmark results provide several levels of comparison for end-users and hardware/software designers, including hardware performance, software utilization, energy consumption. In the future, we will extend our evaluation to more AI accelerators such as FPGAs and more training categories. Another direction is to benchmark the performance and energy efficiency of deep learning inference tasks on both server devices and edge/mobile devices. 

\section*{Acknowledgements} \label{ack}
The research was supported by Hong Kong RGC GRF grant HKBU 12200418. We gratefully acknowledge the support of NVIDIA Corporation with the donation of the Titan X (Pascal) used for this research. We also gratefully acknowledge Google for providing TPUs to support our research in the TensorFlow Research Cloud (TFRC) program.

\bibliographystyle{unsrt}  
\bibliography{cites} 

\end{document}